%% file: ms.tex
\documentclass[12pt,preprint]{aastex}
\usepackage{natbib}
\usepackage{longtable}

\def\timedelay{152.2^{+2.8}_{-3.0}~(1\sigma){~\rm days}}

\begin{document}
\title{
Mid-IR Observations and a Revised Time Delay for the Gravitational
Lens System Quasar HE 1104--1805
}

\author{Shawn Poindexter\altaffilmark{1},
	Nicholas Morgan\altaffilmark{1},
	Christopher S. Kochanek\altaffilmark{1},
	Emilio E. Falco\altaffilmark{2}
}
\altaffiltext{1}{Department of Astronomy, Ohio State University, 
140 West 18th Avenue, Columbus, OH 43210, USA,
(sdp,nmorgan,ckochanek)@astronomy.ohio-state.edu}
\altaffiltext{2}{Harvard-Smithsonian Center for Astrophysics,
60 Garden St., Cambridge, MA 02138, USA}

\begin{abstract}

The mid-IR flux ratios $F_{\rm A}/F_{\rm B} = 2.84\pm0.06$
of the two images of the gravitationally lensed quasar HE 1104--1805
show no wavelength dependence to within $3\%$ across $3.6-8.0~\mu{\rm m}$,
no time dependence over 6 months and agree with the broad emission
line flux ratios.
This indicates that the mid-IR emission likely comes from scales large
enough to be little affected by microlensing and that there
is little differential extinction between the images.
We measure a revised time-delay between these two images of
$\timedelay$ from R and V-band data covering 1997 to 2006.
This time-delay indicates that the lens has an approximately flat rotation
curve over scales of 1-2 $R_e$.
We also observed uncorrelated variations of $\sim$$0.05~\rm mag~yr^{-1}$ which we
attribute to microlensing of the optical emission from the accretion disk.
The optical colors have also changed significantly in the sense that
image A is now redder than image B, rather than bluer as it was in 1993.

\end{abstract}


\section{Introduction
\label{sec:intro}}

The fluxes of gravitationally lensed images are affected by absorption in the
interstellar medium
(ISM, e.g. \citealt{Falco99}) and magnification perturbations from satellites
(substructure, \citealt{Dalal02}) and
stars (microlensing; for a review see \citealt{Wambsganss06}).
The simplest means of disentangling these effects
is to explore the changes in the flux ratios with wavelength and time.
In particular,
the mid-IR flux ratios should be immune to microlensing,
because of the large size of the mid-IR emission region (e.g. \citealt{Barvainis87})
and insensitive to dust extinction because of the long wavelength.
Broad emission line regions should also be insensitive to microlensing because of
their large size (but see \citealt{Richards04}),
but may require corrections
for extinction (e.g. \citealt{Munoz04}).
Neither the mid-IR nor the broad line flux ratios are immune to
magnification perturbations from substructure.
Unfortunately, it is difficult to measure mid-IR flux ratios of lenses
because of the low sensitivity of even the best ground-based telescopes
and the low resolution of current space-based mid-IR telescopes.
Nonetheless, slow progress has been made with the measurement of mid-IR flux
ratios for QSO 2237+0305 \citep{Agol00} and
for PG1115+080 and B1422+231 \citep{Chiba05}.
Since the mid-IR probably provides the best means of isolating
the effects of substructure from those of microlensing and the ISM,
we have tried to obtain Spitzer IRAC mid-IR observations when
feasible to resolve the lens system and measure the flux ratios.

Time-delay measurements have traditionally been pursued for
determining the Hubble constant ($H_0$) independently
of local distance indicators \citep{Refsdal64}.
Alternatively, one can assume a value of $H_0$ and use the time-delay information to
learn about the mass profile of the lens (e.g. \citealt{Kochanek02}).
Time-delay measurements are also important for distinguishing intrinsic
source variability from microlensing variability caused by stars
in the lens galaxy.
Once the two phenomena are separated, 
analysis of the microlensing variability can yield estimates
of the mean stellar mass and the stellar surface density of
the lens galaxy and be used to measure the size
of the source quasar (e.g. \citealt{Kochanek07}).
The level of microlensing variability in the optical is thus complementary to
the mid-IR as a probe of magnification perturbations in
gravitational lenses.

Here we report the results of new observations of
HE 1104--1805, a doubly imaged radio-quiet quasar at $z_s=2.319$
with a separation of $3\farcs15$ from the Hamburg/ESO survey \citep{Wisotzki93}.
The lens at $z_l = 0.729$ was discovered in the near-IR by \citet{Courbin98}
and with {\it HST} \citep{Remy98,Lehar00}.
After a series of initial attempts at measuring a time-delay
\citep{Wisotzki98,Gil-Merino02,Schechter03},
\citet{Ofek03} succeeded with an estimate of 
$161^{+7,+34}_{-7,-11}$ ($1\sigma, 2\sigma$) days that was
confirmed by \citet{Wyrzykowski03}.
One source of difficulty is that the lens suffers considerably
from microlensing effects.  In particular, the flux ratios of the
broad emission lines and the continuum
are dramatically different \citep{Wisotzki93}, the
continuum flux ratio varies with wavelength
\citep{Wisotzki98} and \citet{Schechter03}
found uncorrelated short time-scale variability between the images
at the level of 0.06 magnitudes.

In this paper we present new photometric data for HE 1104--1805 from the
Spitzer Space Telescope (SST), 
the Small and Moderate Aperture Research Telescope System (SMARTS) and
the Southern Astrophysical Research (SOAR) telescope in \S\ref{sec:data}.
We combine our new R-band data with the data from \citet{Schechter03},
\citet{Ofek03}, and \citet{Wyrzykowski03} to make an
improved time-delay estimate for HE 1104--1805 in \S\ref{sec:timedelay}
and we interpret its consequences for the lens model in \S\ref{sec:lensmodel}.
We summarize our results in \S\ref{sec:conclusions}.

\section{Data
\label{sec:data}}


In this section we discuss our mid-IR, near-IR and optical observations.
The mid-IR photometry is presented in Table \ref{tab:IRACdata} and
the optical and near-IR photometry is presented in Table \ref{tab:lcurve}.
Each of these observations make use of the {\it HST} astrometry for
the lens and image positions and {\it HST} photometry for the lens galaxy profile.

\subsection{{\it HST} Observations\label{sec:HST}}

HE 1104--1805 was originally observed with {\it HST} by \citet{Lehar00},
but we obtained deeper NICMOS NIC2 F160W (H-Band) observations as part
of GO-9375 on 17 December 2003 (HJD 2452990.9) with the aim of improving the
lens galaxy astrometry and profile and studying the quasar host galaxy.
These covered 5 orbits
and included observations of
a nearby PSF template star.  The observations of the lens consisted
of 22 dithered sub-exposures with 17, 1, 3, and 1 sub-exposures
having integration times of 640, 448, 128 and 64 seconds respectively
for a total exposure time of 3.27 hours.  The short exposures were
simply used
to fill orbits given the timing restrictions of the long exposures.  The
images were analyzed
as in \citet{Lehar00}.  While the relative astrometry of the
quasar images is identical to the single-orbit
results in \citet{Lehar00}, the new estimate of the
position of the lens galaxy relative to A
($\Delta\hbox{RA},\Delta\hbox{Dec}=0\farcs965\pm0\farcs003$,$-0\farcs500
\pm0\farcs003$)
is shifted by approximately $0\farcs01$ ($2\sigma$) in both coordinates.
The quasar magnitudes are $15.57\pm0.03$ and $17.04 \pm 0.03$
for images A and B, respectively.  The lens galaxy is estimated to have a major axis
effective
radius of $R_e = 0\farcs72 \pm 0\farcs07$, an axis ratio
$q=0.80 \pm 0.01$, a major axis position angle $48^\circ \pm 4^\circ$
and a total magnitude $\hbox{H}=17.52\pm0.09$~mag.  These
parameters for the lens galaxy are all consistent with the
earlier results in \citet{Lehar00}.

\subsection{Spitzer Space Telescope IRAC Observations
\label{sec:IRAC}}

We observed HE 1104--1801 with SST and the Infrared Array Camera (IRAC; 
\citealt{Fazio04}) on
15 June 2005 and 2 January 2006 as part of Spitzer program 20451.
Each observation consisted
of 36 dithered 10.4 second images in each of the four IRAC channels: 
$3.6\mu{\rm m}, 4.5\mu{\rm m}, 5.8\mu{\rm m},$ and $8.0\mu{\rm m}$.
The FWHM of the PSF at $3.6\mu{\rm m}$ is $\sim$$1\farcs5$, so we
can easily resolve the QSO images, but we must worry
about the flux of the lens galaxy that is only $1\farcs0$ from the brighter QSO image.
The spectral energy distribution of the lens galaxy will peak near 2.7 $\mu$m
and then steadily drop as we progress through the IRAC bands.
Thus, as the resolution of the observations diminishes the emission from
the galaxy also becomes less important.
We drizzled the images using the MOPEX
package\footnote{\url{http://ssc.spitzer.caltech.edu/postbcd/}}
to a $0\farcs3$ per pixel scale.
To reduce the number of free parameters, we fix the relative positions
of the QSO images and the lens galaxy as well as the structure of the
lens galaxy using the {\it HST} observations of \S\ref{sec:HST}.
We used the methods of \citet{Lehar00}
and the point response functions (PRF) from the MOPEX package to measure the
fluxes of the three components.
The PRFs are 2x oversampled relative to our drizzled images,
so we binned them to match our images.

We used a bootstrap resampling method to estimate the errors
in the fluxes and flux ratios.
For each trial we randomly chose $36$ images with replacement from the $36$
sub-images, drizzled these images together and
then we fitted this synthetic image to estimate the component fluxes.
We repeated this $200$ times for each of the IRAC bands, and then used the
standard deviation of the flux ratios measured from the synthetic images
as an estimate of the errors on the measured flux ratios.
This process should naturally include both Poisson and systematic errors due
to image pixel sampling in our uncertainties.
The errors for the individual image magnitudes and the lens
galaxy magnitude were estimated in the same way.
The quasar flux ratios are the same in both epochs and in all four IRAC channels
(see Table \ref{tab:IRACdata}), and they agree with
the emission line flux ratios of $2.8$ (with no error given) measured by \citet{Wisotzki93}.
The galaxy magnitudes are given in Table \ref{tab:IRACdata}, but the
uncertainties are too large to measure the colors accurately because
of flux trade-offs with the brighter quasar image.
Based on the lens redshift, an early-type galaxy should have
IRAC colors of $[3.6]-[4.5] = -0.05$ and
$[5.8]-[8.0] = -0.03$ (R. Assef, private communication 2006).
Our measurements are consistent with these colors with large uncertainties.

\subsection{Monitoring Data
\label{sec:monitoringdata}}

Most of our optical and near-IR data
was obtained using the queue-scheduled SMARTS
$1.3$m telescope with the ANDICAM optical/infrared camera \citep{DePoy03}.
Between $3$ December $2003$ until
$22$ May $2006$ we obtained 97 epochs in our primary monitoring band,
the R-band, and sparse
observations in the B and I-bands.
A simultaneous J-band image is obtained at all epochs.
We used the method of \citet{Kochanek06a} to obtain the relative photometry of both images.
For each band we determined the flux of the lens galaxy by fitting the 
images as a function of galaxy flux and then adopted the flux which optimized
the goodness of fit for all images simultaneously.
For calibration in future epochs, we
list the positions and relative fluxes for our field reference stars in
Table \ref{tab:rel_astro}.  The relative magnitudes for images A and B
given in Table \ref{tab:lcurve} are
with respect to the first star in Table \ref{tab:rel_astro}.

\subsection{Near-IR SOAR Data
\label{sec:soardata}}

We obtained a single epoch of near-IR data for HE 1104--1805 using the SOAR 4.1m
telescope at Cerro Pachon, Chile.  Observations were taken on 14 January
2006 with OSIRIS \citep{DePoy93} operating at f/7,
giving a plate scale of 0.139
arcsec/pixel and a field of view of 2.4 arcmin on a side.  The exposures
consisted of 16, 18 and 17 box-dithered images through JHKs Barr filters,
respectively, with exposure times of 10 seconds per image.  Along with the
$0\farcs7$--$0\farcs75$ FWHM seeing, this provided sufficient signal to noise in all
filters to identify the two quasars and the lensing galaxy by eye after
proper sky subtraction and image registration.  Flux from image B was well
isolated, but the 5 pixel seeing disk and 8 pixel image A-lens galaxy
separation meant that significant flux overlap between these two
components was unavoidable.  We used the same technique to optimize the
galaxy flux as with the SMARTS data, namely choosing the galaxy flux that
gave the lowest $\chi^2$ residuals after again fixing the galaxy size to
that measured from the {\it HST} data,
which for this single-epoch data
simply meant fitting all three component fluxes simultaneously while again
holding the relative positions fixed.  As a check on the
flux decomposition, the SOAR J-band A-B magnitude difference of $-1.33\pm
0.10$ mag is consistent with the SMARTS J-band value of $-1.39\pm0.03$ mag
taken a night earlier.
The fitting results for all three filters are given in Table \ref{tab:lcurve}.

\section{Time Delay Estimates and Microlensing
\label{sec:timedelay}}

\begin{figure}
\plotone{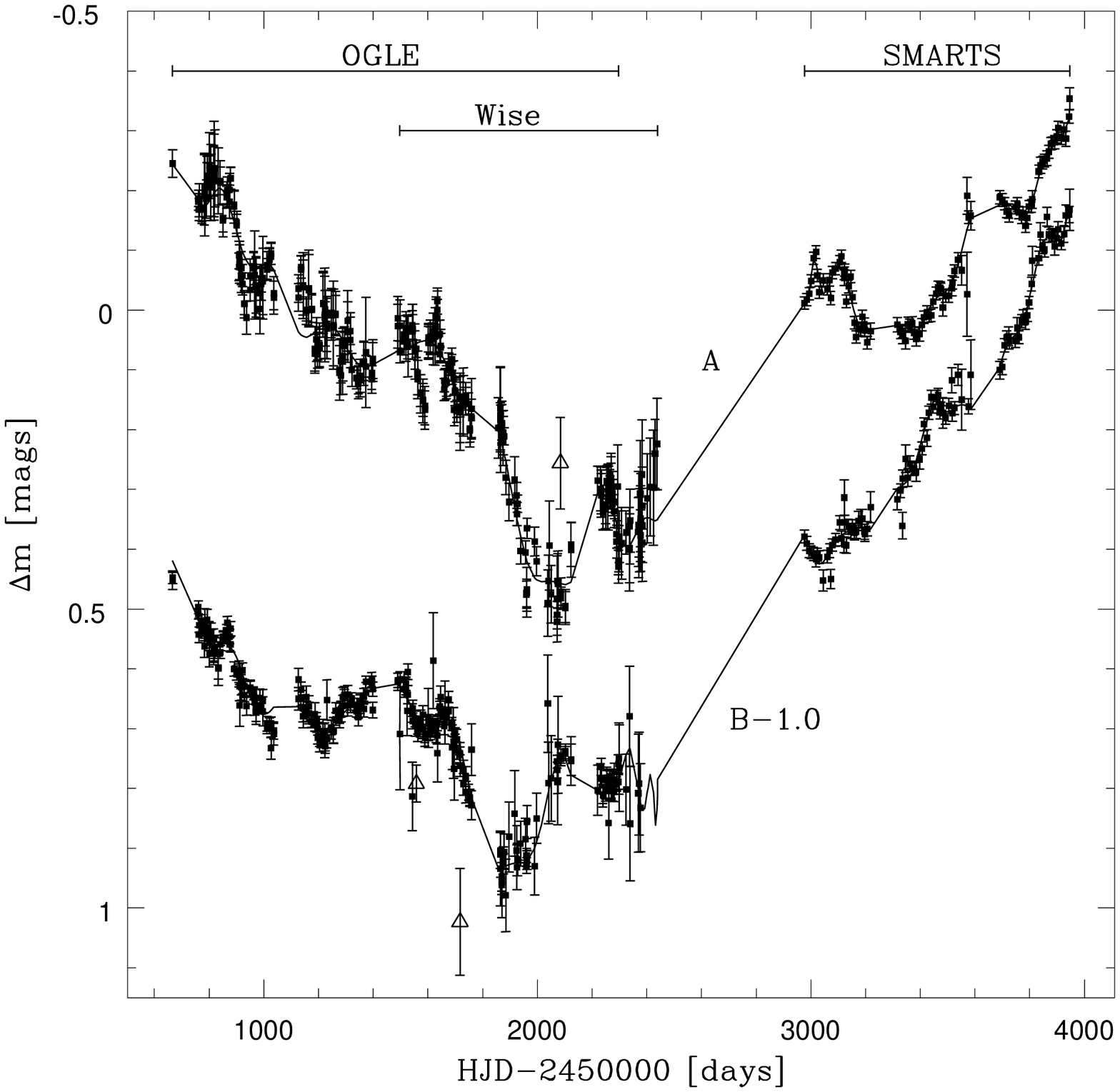}
\caption{The light curve used for our time-delay estimate.  It includes only the
SMARTS R-band, OGLE V-band, and Wise R-band data, where bands at the top
indicate the period during which each observatory monitored the system.
The three open symbols indicate
Wise data points that were not considered in the final time-delay estimate.
The (lower) image B light curve has been shifted by one magnitude.
Points with errors greater than 0.1 magnitudes have been suppressed for
clarity.
\label{fig:lc}}
\end{figure}

For the time-delay determination we combined the SMARTS R-band data 
with the Wise Observatory R-band data from \citet{Ofek03} and
the Optical Gravitational Lensing Experiment (OGLE) V-band
data \citep{Schechter03,Wyrzykowski03}.  The V-band data was adjusted to match
the R-band data as in \citet{Ofek03}.  The resulting light curve is shown in
Figure \ref{fig:lc}.
Three Wise observations of image A (MJD epochs 2084.254, 1717.253,
and 1557.520) were clearly outliers from the light curves and they
were masked in our time-delay estimate.  
In total we used 408 epochs (or 816 data points) to estimate
the time delay.

In the OGLE data, 
the A image shows rapid variability at the level of 0.06 mag
on one month timescales that is not observed in image B, which \citet{Schechter03}
argued this is due to the microlensing of hot spots in the accretion disk.
In order to compensate for this systematic noise,
we attempted to rescale the errors for each
image and for each observatory in an unbiased way
by fitting a Legendre polynomial with $N_s = 40$ to
each image separately with no time delay to estimate the factor
by which the error estimates for each data set must be rescaled
to be consistent with a smooth fitting curve.
The factors for rescaling the Wise, OGLE, and SMARTS data for
image A are 2.4, 4.6, and 1.0, and those
for image B are 1.1, 1.2, and 0.9.
Since using a polynomial order of $N_s = 40$ for the
error renormalization is somewhat arbitrary,
we also tried $N_s = 70$ and the time final time-delay estimate
changed by only $\sim$$1\%$.

We determined the time delay using the polynomial fitting methods of
\citet{Kochanek06a}.  We divided the data into two observing periods at
JD $= 2452708$ because of a considerable gap in the data.
In each observing period we modeled the flux of the
source by Legendre polynomials of order
$N_s = 20,30,...,80$ and the microlensing variability by Legendre
polynomials of order $N_\mu = 1,2,...,6$.
Each model $m$ then leads to a $\chi^2$ fit statistic $\chi^2_m(\Delta t)$
that can be used to estimate the delay and whose value at the
minimum relative to the number of degrees of freedom $N_{\rm dof}(m)$ can
be used to evaluate the significance of the fit.
One issue with the polynomial method is the choice of polynomial order.
Our previous approach \citep{Kochanek06a} used the F-test to determine when
increasing the order of the fitting polynomials
was statistically unnecessary.
Using the same technique for HE 1104--1805
we find that orders more complex than $N_s = 50, N_\mu = 4$ are
not required by the data and this model gives a time-delay estimate of
$157.2\pm2.6$ days.

A potential issue with using the F-test is that the restriction of
the final estimate to a single model may underestimate the uncertainties
by neglecting the effects of shifts in the delay due to changes
in the source or microlensing model.  We can use Bayesian methods
to perform a reasonable average over the results for different models by
adding an information criterion for the significance of changes in the
numbers of model parameters.
If model $m$ has $N_m$ parameters, then we assign it a likelihood of
fitting the data of
\begin{equation}
P(D|\Delta t, m) \propto \exp(-\chi^2_m(\Delta t)/2 - k N_m),
\label{eqn:likelihood}
\end{equation}
so a model with more parameters needs to reduce $\chi^2_m$ by $2k$
for each additional parameter to represent an improvement in the fit.
We considered the two cases of $k=1$ (which corresponds to the Akaike
Information Criterion or AIC) and $k=\ln n$ (where $n=$ number of
data, which corresponds to the Bayesian Information Criterion or BIC).
The AIC treats new parameters more favorably than the BIC method.
The best models using the AIC weights
have $N_s = 80, N_\mu = 5$, while those for the
BIC weights have $N_s = 30, N_\mu = 4$.
The time-delay estimate is then the average over all the models
\begin{equation}
P(\Delta t | D) \propto \sum_{m} P(D | \Delta t, m),
\label{eqn:avglikelihood}
\end{equation}
where we assume a uniform prior for the different models after adding the
information criterion.
Thus, models of similar likelihood contribute equally to the delay estimate,
while models with low likelihood contribute little.

\begin{figure}
\plotone{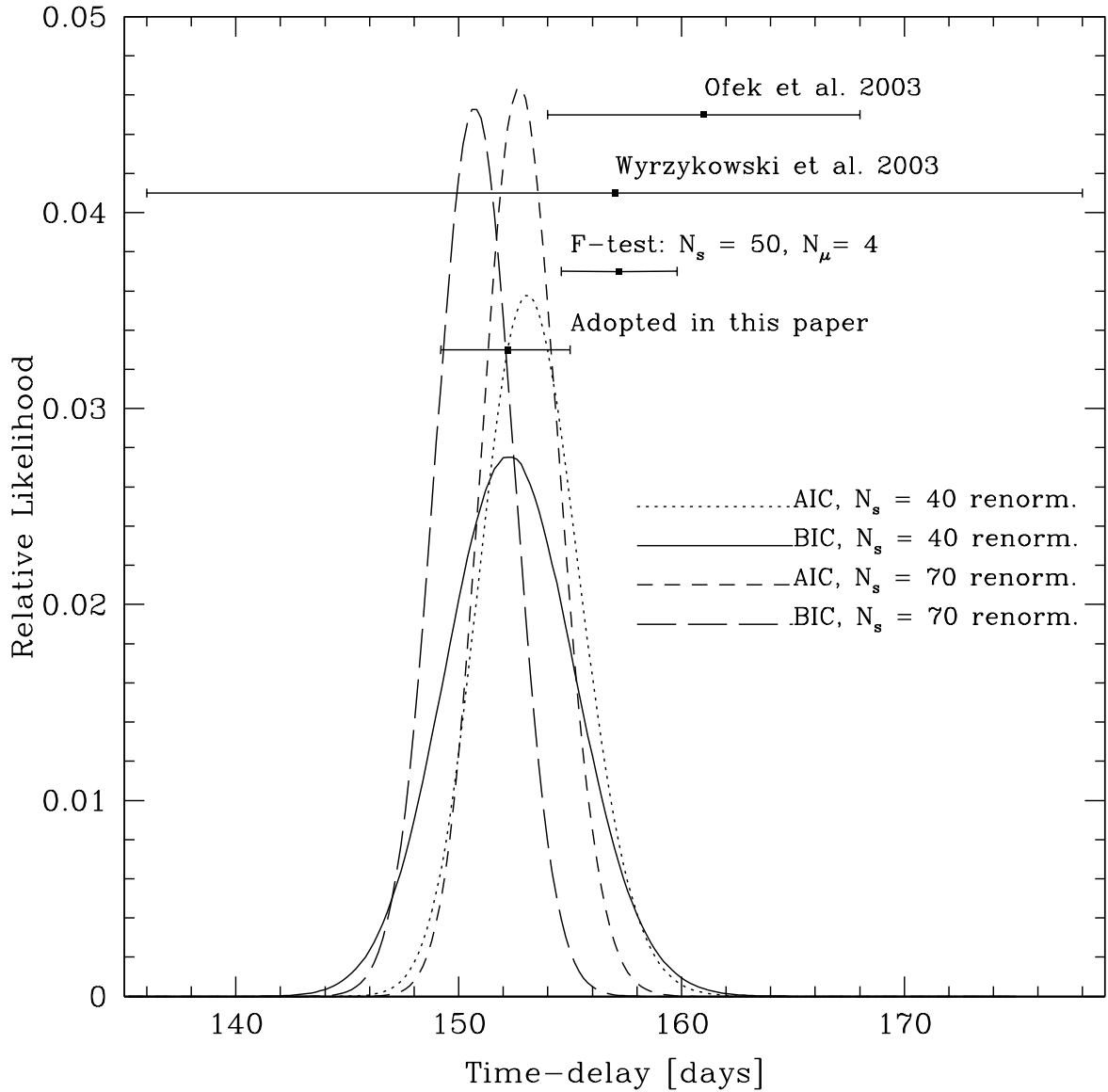}
\caption{
The relative likelihood for each time-delay $\Delta t$ for the
two different error renormalization schemes ($N_s = 40$ or $N_s = 70$)
and the two different information criterion (AIC and BIC).
The points with error bars indicate our adopted and 
the Ofek \& Maoz (2003, excluding
systematic errors), OGLE and F-test delay estimates.
We adopt the broadest of the Bayesian estimates (BIC $N_s = 40$)
as our standard estimate.
\label{fig:timedelay}}
\end{figure}

Figure \ref{fig:timedelay} shows the resulting probability distribution for both of
these criteria and with the photometric uncertainties renormalized using
either the $N_s = 40$ or $N_\mu = 70$ model for a smooth light curve.
The result for these four cases
(Table \ref{tab:timedelays}) are mutually consistent and roughly consistent with our previous
F-test approach.
We adopt the broadest of these four Bayesian estimates,
$\Delta t_{\rm AB} = t_{\rm A} - t_{\rm B} = \timedelay$
in the sense that image B leads image A as our standard estimate.
None of the Bayesian models gives significant weight to the model singled
out by the F-test, although they are marginally consistent and
the difference in the delay is only $3\%$.
The delay in the model favored by the F-test is curiously shifted from
the time-delay given by most models with comparable polynomial order.
It also has a $\chi^2(\Delta t)$ minimum
intermediate to those favored by both the AIC and BIC criterion,
so it is given little weight in either case.
The Bayesian estimates are consistent with the
best previous measurements of 
$161^{+7,+34}_{-7,-11}$ ($1\sigma, 2\sigma$) days by \citet{Ofek03} and
$157\pm21$ days by \citet{Wyrzykowski03}.
These analysis-dependent shifts suggest that the measurement uncertainties
should be interpreted conservatively.  Moreover, the formal $2\%$ measurement
errors are in the regime where the cosmic variance of
order $5\%$ in the delays due to the structures
along the line of sight 
are a significant source of uncertainty (e.g. \citealt{Barkana96}).

\begin{figure}
\plotone{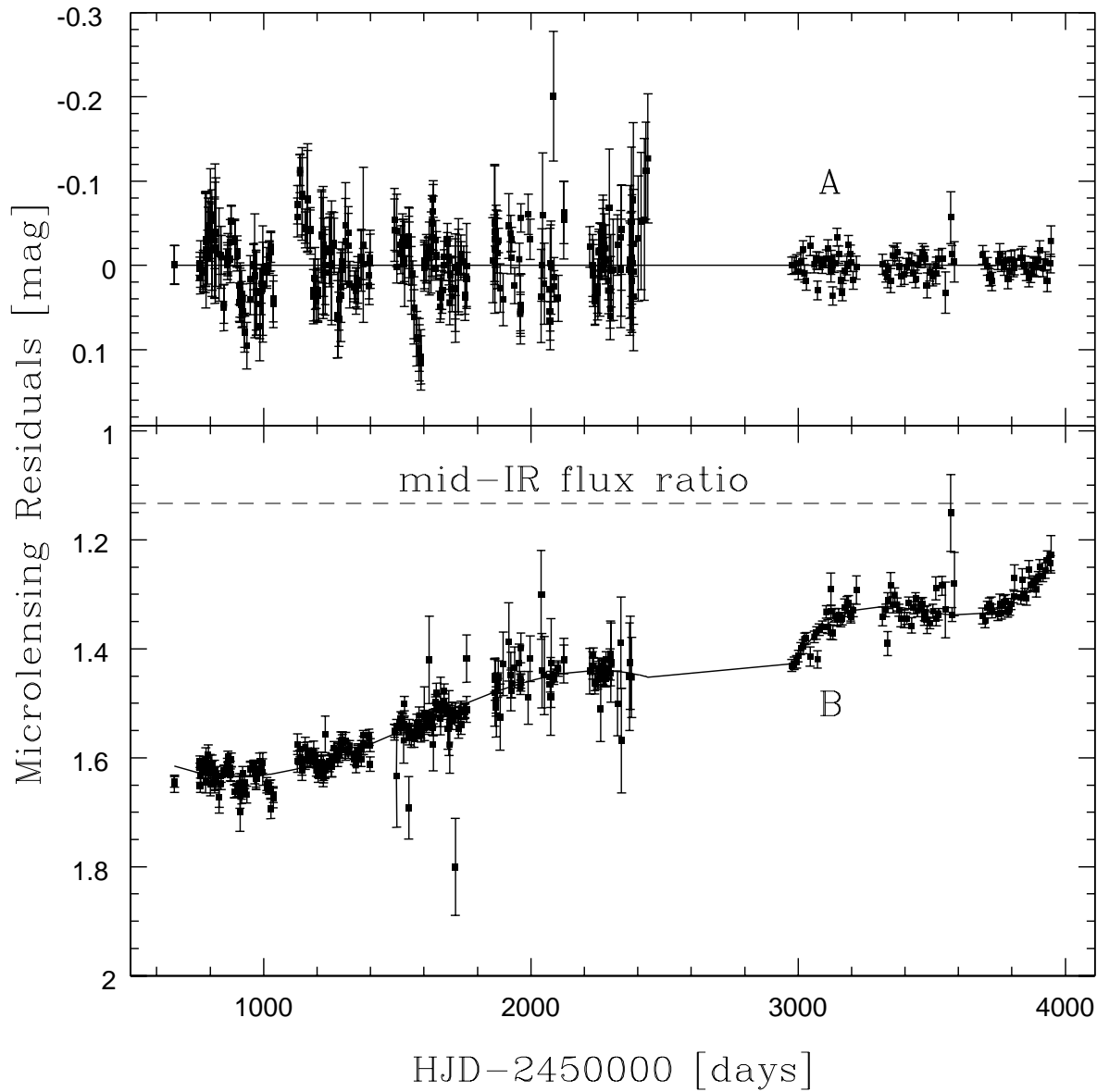}
\caption{
Microlensing variability in HE 1104--1805.
The points show the data after subtracting the model for the source
variability in our standard model, and the curves show the model for the
microlensing variability.
The fit is broken at HJD $= 2452708$ into two segments.
By definition we assign the variability to
B, but we can only determine the difference between A and B.
Note how the flux ratio has been steadily approaching that measured from the
broad emission lines or at mid-IR wavelengths.
\label{fig:micro}}
\end{figure}

\begin{figure}
\plotone{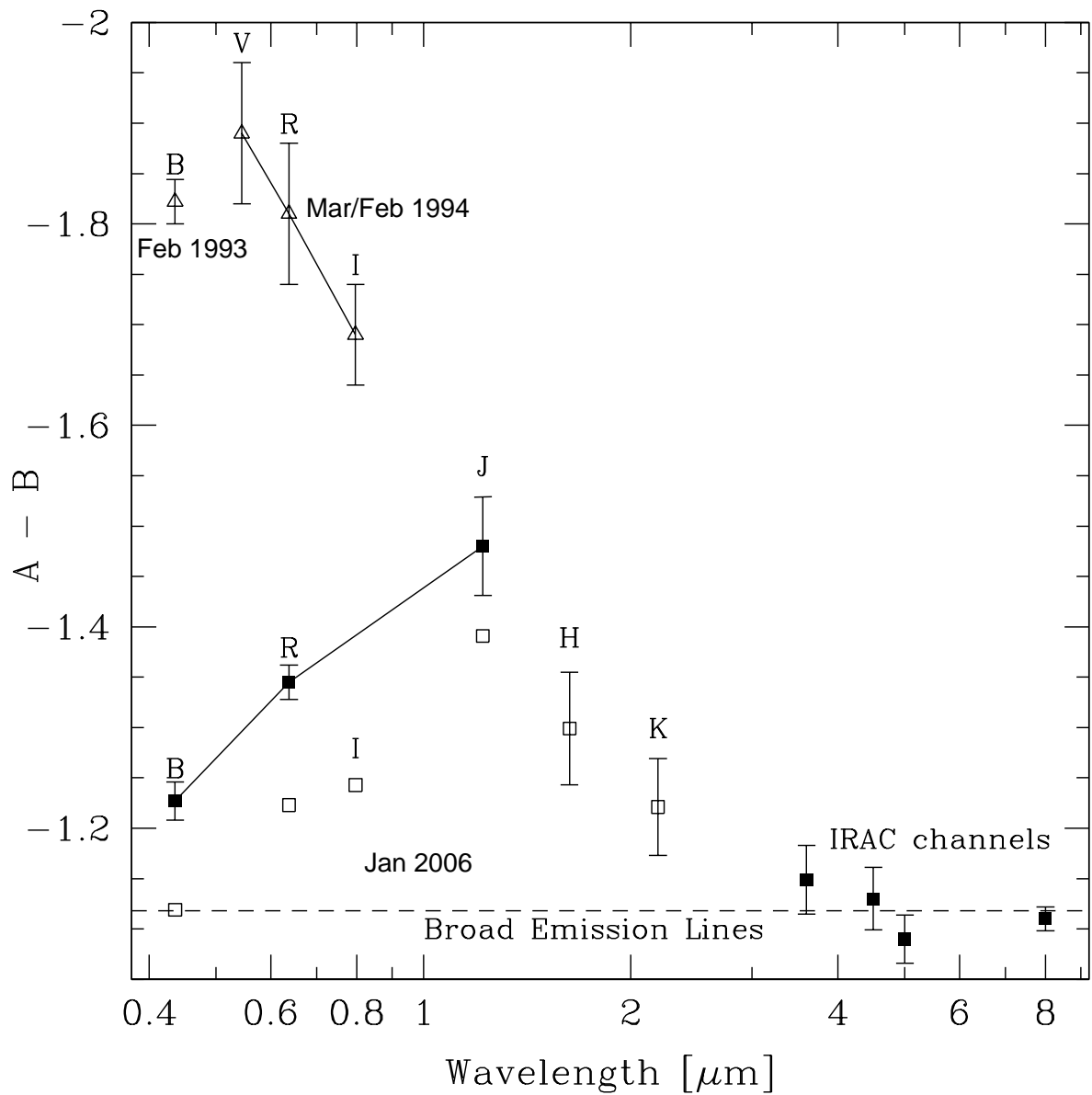}
\caption{
Flux ratio versus wavelength.
Solid (open) symbols represent flux ratios that are (are not) corrected
for the time delay.
Squares are used for the January 2006 data while triangles are used for
flux ratios from 1993/1994.
Note that the recent flux ratios are shifted by about 0.1 magnitudes
when we correct for the time delay due to the relatively rapid source variability.
While we do not have time-delay
corrected H and K-band flux ratios we expect they are affected less by the
time delay because the IRAC channel flux ratios did not change
on the time delay scale.
Note how the
flux ratio is now much closer to the mid-IR and emission line
flux ratios and that the optical flux ratios now have A redder than B
while the reverse still holds in the near-IR.
\label{fig:wavelength}}
\end{figure}

Given the time delay, we can shift the light curves to obtain an
estimate of the fluctuations due to
microlensing, as shown in Figure \ref{fig:micro} for our standard model.
Over the decade since its discovery the flux ratio has been changing
by approximately $0.05 {\rm~mag~yr^{-1}}$ with considerably more
curvature over the last few years as the flux ratio approaches that
measured in the mid-IR or from the broad emission lines.
The long time delay means that there is significant microlensing
variability on the time scale of the delay, making it essential to
explore a broad range of microlensing models when estimating the delay.
Figure \ref{fig:wavelength} shows the wavelength-dependent flux ratios
of the images at the time of the Spitzer observations and when it was
discovered \citep{Gil-Merino02,Remy98}.
The optical A$-$B colors have become significantly redder and the flux ratio has
become much smaller since the discovery of this system.
There is also a curious reversal in the color trends near $1.2 \mu{\rm m}$.

\section{Lens Models
\label{sec:lensmodel}}

We modeled the system by combining
a constant $M/L$ de Vaucouleurs model with an NFW halo
using the {\tt lensmodel} program
of the {\tt gravlens} package \citep{Keeton01}.
The constant $M/L$ model represents the visible lens galaxy which we
model with ellipticity of $e=1-q=0.20\pm0.04$,
a position angle of $PA = 48\pm4\arcdeg$ (East through North),
and an effective radius of $R_e = 0\farcs72$ as determined from {\it HST}
observations discussed in \S\ref{sec:HST}.
We broadened the uncertainties in the axis ratio from the fits in \S\ref{sec:HST}
since we also use this axis ratio and position angle for the
ellipsoidal NFW component.
We used a prior on the external shear of $\gamma = 0.05\pm0.05$,
and we assumed a cosmology with $\Omega_m = 0.3, \Omega_\Lambda = 0.7$, and
$H_0 = 72\pm7 \rm~km s^{-1} Mpc^{-1}$ \citep{Freedman01}.
We used a scale length of $r_s = 10\farcs0$ for the NFW model.
The results will depend only weakly on $r_s$ if it is significantly
larger than the image radii.
We constrained the flux ratios to $F_{\rm A/B} = 2.84\pm0.06$,
the average of the mid-IR flux ratios.
As a two image lens with modest image magnifications ($11.5$ and $4.0$)
the HE 1104 image fluxes should be relatively immune to perturbations
from substructure (e.g. \citealt{Mao98}).
We used our newly estimated time delay
$\Delta t_{\rm AB} = t_{\rm A} - t_{\rm B} = 152\pm9$ days
with the errors broadened to $6\%$ to account for cosmic variance.

\begin{figure}
\plotone{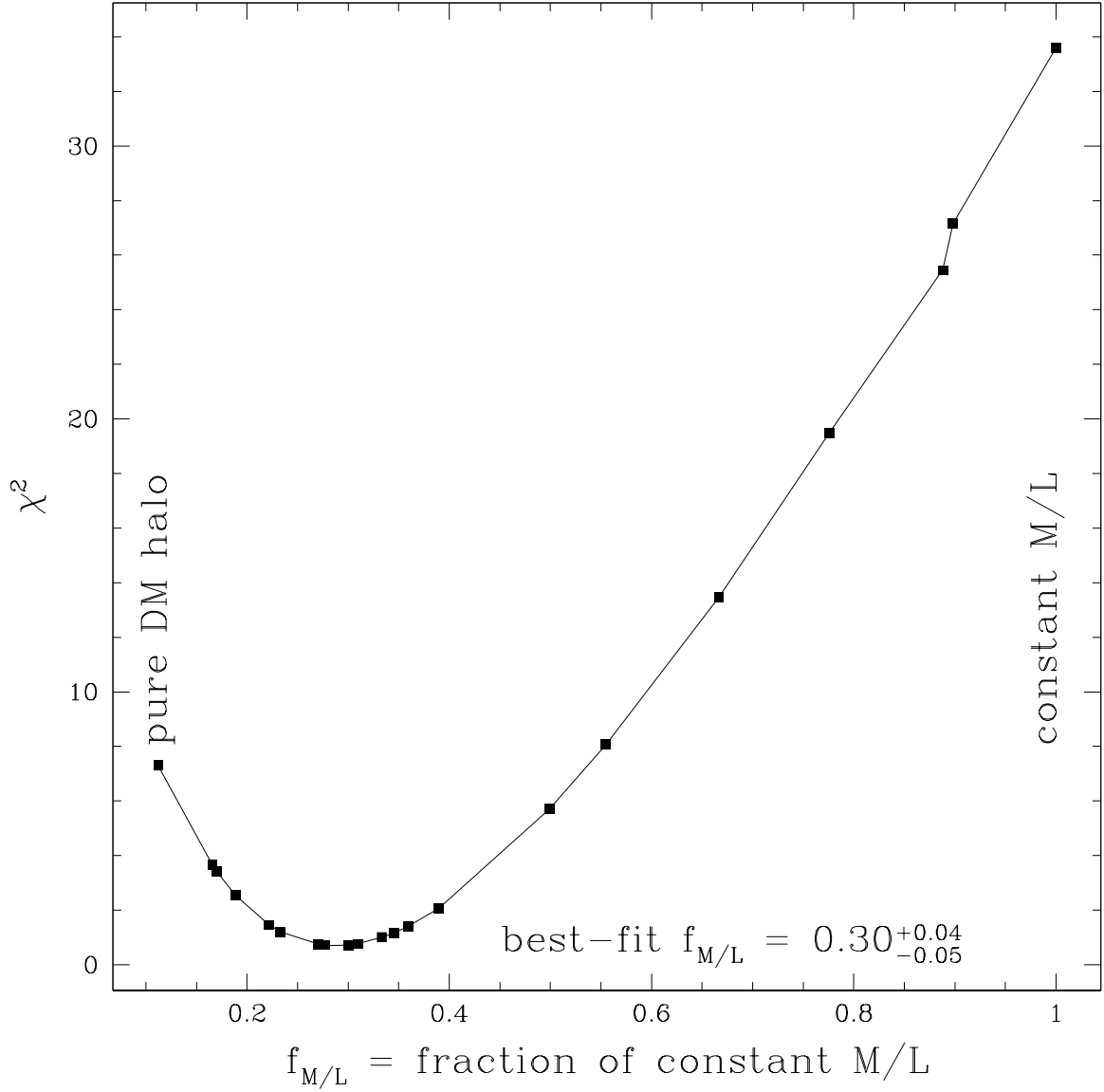}
\caption{
Goodness of fit for the 
``galaxy plus halo'' models.
The $\chi^2$ is plotted as a function of $f_{M/L}$, which is the
mass of the de Vaucouleurs component divided by the mass of the
best-fitting constant-$M/L$ de Vaucouleurs model.
\label{fig:chi2fraction}}
\end{figure}

\begin{figure}
\plotone{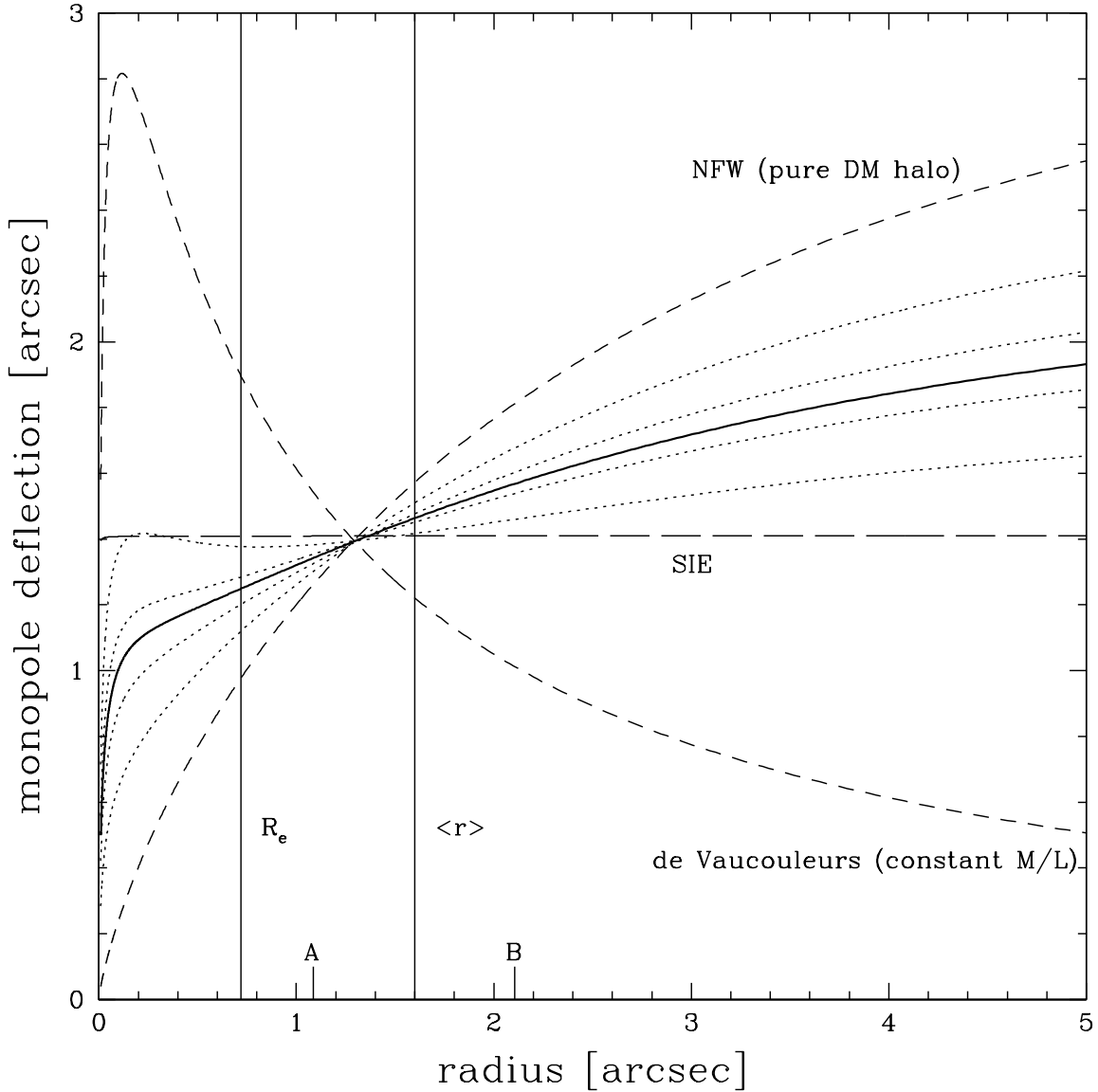}
\caption{Radial deflection profiles, which roughly correspond to the
square of the rotation curve,
for the de Vaucouleurs plus NFW models.
The solid curve is the best-fitting model with $f_{M/L} = 0.30$.
The four dotted curves are the 1 and 2 $\sigma$ bounds
to the best-fit model.
The dashed lines show the profile for the pure NFW model,
a singular isothermal ellipsoid (SIE) and a constant $M/L$ model.
The vertical lines mark the effective radius $R_e$ and the mean radius
of the images $\langle r\rangle$. The radial distances of the images from the
center of the galaxy are labeled ``A'' and ``B''.
\label{fig:monopole}}
\end{figure}

Figure \ref{fig:chi2fraction} shows the goodness of fit as a
function of $f_{M/L}$ where $f_{M/L}$ is the fraction of mass in the
de Vaucouleurs component compared to a constant $M/L$ model
in which the NFW component has no mass ($f_{M/L} = 1$).
The best-fit model has $f_{M/L} = 0.30^{+0.04}_{-0.05}$.
Figure \ref{fig:monopole} shows the monopole deflection profile of these models,
which is similar to the square of the rotation curve.
We see that the best-fit models have quasi-flat rotation curves
on scales of $1-2 R_e$.  In
fact a model using a singular isothermal ellipsoid fits
the data reasonably well, but is formally ruled out at 
approximately $95\%$ confidence ($\Delta\chi^2 = 3.6$).

\section{Conclusions
\label{sec:conclusions}}

We find that the mid-IR flux ratios of $F_{\rm A}/F_{\rm B} = 2.84\pm0.06$
in HE 1104--1805
agree with the broad emission line flux ratios of 2.8 \citep{Wisotzki93}.
The flux ratios from $3.6-8.0~\mu\rm m$ did not change over the 6 month
period between the two IRAC epochs.
The optical flux ratios have fallen from
4.6 when the lens was discovered to a value of 3.1 today that
approaches the ratio in the mid-IR and emission lines,
indicating that there is significant microlensing of the quasar accretion disk.
This means, as expected, that the mid-IR/broad line emission regions are large
compared to the Einstein radius of the microlenses
$R_{\rm E} = 3.6\times10^{16}~{\rm cm}~(\left<M\right>/h M_\sun)^{1/2}$,
while the optical emission regions are significantly more compact.
The microlensing is clearly chromatic, as image A is now significantly
redder relative to B than when the system was discovered in 1993.
It also seems that as A has faded relative to B, the excess variability
in A observed by \citet{Schechter03} has vanished.
We will explore this quantitatively in \citet{Poindexter07}.

We have also determined a new time delay for HE 1104--1805,
$\Delta t = t_{\rm A} - t_{\rm B} = \timedelay$ in the sense that
image B leads image A.
This uncertainty should be interpreted conservatively because there are
some variations in the estimate with the analysis method and because
cosmic variance is a significant contribution to the
uncertainties when the delay is used to model the mass
distribution of the lens.
If we model the lens galaxy mass distribution as the observed
de Vaucouleurs profile embedded in an NFW halo,
we find the best-fit lens galaxy
model has a stellar mass fraction in the de Vaucouleurs component that is
$f_{M/L} = 0.30^{+0.04}_{-0.05}$ of a constant $M/L$ model.
This ratio corresponds to a deflection profile that is marginally
consistent ($\Delta\chi^2 = 3.56$) with the system 
having a flat rotation curve.
Thus HE 1104--1805 has a mass distribution typical of gravitation lens galaxies
\citep{Rusin05,Treu06}.

\acknowledgments

This work is based in part on observations made with the Spitzer Space Telescope,
which is operated by the Jet Propulsion Laboratory, California Institute of
Technology under a contract with NASA.
Support for this work was provided by NASA through an award
GRT00003172 issued by JPL/Caltech.
Based on observations made with the NASA/ESA Hubble Space Telescope, obtained
at the Space Telescope Institute. STScI is operated by
the association of Universities for Research in Astronomy, Inc. under the NASA
contract NAS 5-26555.
OSIRIS is a collaborative project between the Ohio State University and Cerro
Tololo Inter-American Observatory (CTIO) and was developed through NSF grants
AST 90-16112 and AST 92-18449. CTIO is part of the National Optical Astronomy
Observatory (NOAO), based in La Serena, Chile. NOAO is operated by the
Association of Universities for Research in Astronomy (AURA), Inc. under
cooperative agreement with the National Science Foundation.
We would like to thank Roberto Assef for providing estimates of the IRAC colors
for the lens galaxy.

\begin{deluxetable}{ccccccc}                                             
\tablecolumns{7}                                                               
\tablewidth{0pc}                                                                
\tablenum{1}                                                                    
\tablecaption{SST IRAC Observations of HE~1104$-$1805\label{tab:IRACdata}}
\tablehead{                                                                     
  \colhead{HJD} &
  \colhead{$\lambda~[\mu{\rm m}]$} & 
  \colhead{A/B} &
  \colhead{A} &
  \colhead{B} &
  \colhead{G} 
}
\startdata
\input tab1.tex
\enddata
\tablewidth{35pc}                                                                
\tablecomments{HJD is the Heliocentric JD minus 2450000.  The $\lambda$ column is the
wavelength in $\mu{\rm m}$.  The A/B column is the flux ratio of image A to image B.
The A, B, and G columns give the measured Vega magnitudes with the errors estimated from
the bootstrap resampling technique.
In the second epoch $8~\mu$m observation, we present an upper limit from the
bootstrap sample since our best fit model assigned it a slightly negative flux.}
\end{deluxetable}

\begin{deluxetable}{crrcc}                                             
\tablecolumns{5}                                                               
\tablewidth{0pc}                                                                
\tablenum{2}                                                                    
\tablecaption{Relative Astrometry and Photometry for
Field Reference Stars\label{tab:rel_astro}}
\tablehead{                                                                     
  \colhead{ID} &
  \colhead{$\Delta \rm RA$ ($\arcsec$)} &
  \colhead{$\Delta \rm Dec$ ($\arcsec$)} &
  \colhead{R-band $\Delta m$ (mag)} &
  \colhead{J-band $\Delta m$ (mag)} 
}
\startdata
\input tab2.tex
\enddata
\tablewidth{35pc}                                                                
\tablecomments{The positions are measured with respect to HE 1104 image A.  The
magnitude differences are for the SMARTS R- and J-band data.  For the SOAR
data, only star 1 (set to unity) was available as a local reference star.}
\end{deluxetable}

\begin{deluxetable}{cccccl}
\tablecolumns{6}                                                               
\tablewidth{0pc}                                                                
\tablenum{3}                                                                    
\tablecaption{Light Curves for HE~1104$-$1805\label{tab:lcurve}}
\tablehead{
       \multicolumn{1}{c}{HJD} &
       \multicolumn{1}{c}{$\chi^2$/dof} &
       \multicolumn{1}{c}{\phantom{++}Comp. A\phantom{++}} &
       \multicolumn{1}{c}{\phantom{++}Comp. B\phantom{++}} &
       \multicolumn{1}{c}{\phantom{++}ref. stars\phantom{++}} &
       \multicolumn{1}{c}{\phantom{++}Telescope\phantom{++}} 
}
\startdata
\input tab3.tex
\enddata
\tablecomments{HJD is the Heliocentric JD minus 2450000.
The A and B columns give the image magnitudes relative to the local reference stars.
The ref. stars column gives the average magnitude differences
of the reference stars relative to the campaign mean.
These values are $\equiv 0$ for the three SOAR data points because we
have only one epoch of data.}
\end{deluxetable}

\begin{deluxetable}{ccc}
\tablecolumns{3}                                                               
\tablewidth{0pc}                                                                
\tablenum{4}                                                                    
\tablecaption{Time-Delay Estimates\label{tab:timedelays}}
\tablehead{
       \multicolumn{1}{c}{Criterion} &
       \multicolumn{1}{c}{Renormalization} &
       \multicolumn{1}{c}{Time-Delay}
}
\startdata
\input tab4.tex
\enddata
\tablecomments{
The Criterion column indicates the statistical method and the Renormalization
column indicates the polynomial order used to renormalize the error bars
(see \S\ref{sec:timedelay}).
}
\end{deluxetable}

\end{document}

%% file: tab1.tex
$3536.73$ & $3.6$ & $2.87\pm0.09$ & $14.03\pm0.04$ & $15.17\pm0.02$ & $15.8\pm0.2$ \\
$3536.73$ & $4.5$ & $2.82\pm0.08$ & $13.30\pm0.05$ & $14.67\pm0.04$ & $15.4\pm0.4$ \\
$3536.73$ & $5.8$ & $2.84\pm0.06$ & $12.13\pm0.02$ & $13.26\pm0.02$ & $16.2\pm1.2$ \\
$3536.73$ & $8.0$ & $2.90\pm0.03$ & $10.85\pm0.02$ & $12.00\pm0.01$ & $17.2\pm5.3$ \\
$3737.85$ & $3.6$ & $2.88\pm0.09$ & $14.03\pm0.04$ & $15.18\pm0.02$ & $15.7\pm0.2$ \\
$3737.85$ & $4.5$ & $2.83\pm0.08$ & $13.27\pm0.05$ & $14.40\pm0.03$ & $16.3\pm0.8$ \\
$3737.85$ & $5.8$ & $2.73\pm0.06$ & $12.16\pm0.02$ & $13.25\pm0.02$ & $16.0\pm1.0$ \\
$3737.85$ & $8.0$ & $2.78\pm0.03$ & $10.89\pm0.02$ & $12.00\pm0.01$ & $>15.4$ \\

%% file: tab2.tex
1 &  $   -3.4$ &  $  -15.1$ &      $\equiv 0$    &    $\equiv 0$ \\
2 &  $   31.8$ &  $   -5.3$ &   $\phantom{-}0.731 \pm 0.002$ &       \nodata \\
3 &  $  -45.7$ &  $   23.8$ &   $\phantom{-}0.604 \pm 0.002$ & $0.653 \pm 0.009$ \\
4 &  $   33.6$ &  $  119.3$ &   $-0.003 \pm 0.003$ &       \nodata \\
5 &  $ -143.9$ &  $  -64.4$ &   $-1.896 \pm 0.028$ &       \nodata \\

%% file: tab3.tex
$2976.806$ &$  0.16$ &$\phantom{-} 0.118\pm 0.031$ &$\phantom{-} 1.663\pm 0.092$ &$\phantom{-} 0.024\pm 0.009$ &SMARTS J \\ 
$2976.807$ &$  2.16$ &$-0.012\pm 0.010$ &$\phantom{-} 1.379\pm 0.012$ &$\phantom{-} 0.043\pm 0.005$ &SMARTS R \\ 
$2985.815$ &$  0.19$ &$\phantom{-} 0.134\pm 0.027$ &$\phantom{-} 1.575\pm 0.073$ &$\phantom{-} 0.027\pm 0.009$ &SMARTS J \\ 
$2985.818$ &$  1.53$ &$-0.017\pm 0.010$ &$\phantom{-} 1.390\pm 0.013$ &$\phantom{-} 0.035\pm 0.005$ &SMARTS R \\ 
$2993.822$ &$  0.10$ &$\phantom{-} 0.114\pm 0.029$ &$\phantom{-} 1.526\pm 0.076$ &$\phantom{-} 0.024\pm 0.009$ &SMARTS J \\ 
$2993.826$ &$  2.76$ &$-0.028\pm 0.010$ &$\phantom{-} 1.403\pm 0.013$ &$\phantom{-} 0.034\pm 0.005$ &SMARTS R \\ 
$3000.817$ &$  0.17$ &$\phantom{-} 0.098\pm 0.024$ &$\phantom{-} 1.630\pm 0.063$ &$\phantom{-} 0.030\pm 0.009$ &SMARTS J \\ 
$3000.820$ &$  3.95$ &$-0.049\pm 0.010$ &$\phantom{-} 1.405\pm 0.012$ &$\phantom{-} 0.048\pm 0.005$ &SMARTS R \\ 
$3009.812$ &$  0.15$ &$\phantom{-} 0.061\pm 0.023$ &$\phantom{-} 1.623\pm 0.063$ &$\phantom{-} 0.028\pm 0.009$ &SMARTS J \\ 
$3009.815$ &$  2.31$ &$-0.087\pm 0.010$ &$\phantom{-} 1.408\pm 0.012$ &$\phantom{-} 0.040\pm 0.005$ &SMARTS R \\ 
$3018.776$ &$  1.47$ &$-0.097\pm 0.011$ &$\phantom{-} 1.417\pm 0.014$ &$\phantom{-} 0.021\pm 0.005$ &SMARTS R \\ 
$3023.757$ &$  2.80$ &$-0.058\pm 0.010$ &$\phantom{-} 1.409\pm 0.012$ &$\phantom{-} 0.047\pm 0.005$ &SMARTS R \\ 
$3030.775$ &$  0.13$ &$\phantom{-} 0.106\pm 0.024$ &$\phantom{-} 1.531\pm 0.061$ &$\phantom{-} 0.032\pm 0.009$ &SMARTS J \\ 
$3030.779$ &$  2.86$ &$-0.030\pm 0.010$ &$\phantom{-} 1.413\pm 0.012$ &$\phantom{-} 0.048\pm 0.005$ &SMARTS R \\ 
$3044.752$ &$  0.12$ &$\phantom{-} 0.124\pm 0.030$ &$\phantom{-} 1.567\pm 0.081$ &$\phantom{-} 0.027\pm 0.009$ &SMARTS J \\ 
$3044.753$ &$  0.83$ &$-0.050\pm 0.011$ &$\phantom{-} 1.452\pm 0.020$ &$-0.026\pm 0.005$ &SMARTS R \\ 
$3060.667$ &$  0.10$ &$\phantom{-} 0.122\pm 0.028$ &$\phantom{-} 1.609\pm 0.077$ &$\phantom{-} 0.024\pm 0.009$ &SMARTS J \\ 
$3060.669$ &$  3.00$ &$-0.035\pm 0.010$ &$\phantom{-} 1.413\pm 0.012$ &$\phantom{-} 0.048\pm 0.005$ &SMARTS R \\ 
$3068.654$ &$  0.14$ &$\phantom{-} 0.117\pm 0.025$ &$\phantom{-} 1.556\pm 0.063$ &$\phantom{-} 0.031\pm 0.009$ &SMARTS J \\ 
$3068.657$ &$  1.88$ &$-0.050\pm 0.010$ &$\phantom{-} 1.403\pm 0.013$ &$\phantom{-} 0.034\pm 0.005$ &SMARTS R \\ 
$3072.717$ &$  2.00$ &$-0.020\pm 0.011$ &$\phantom{-} 1.450\pm 0.018$ &$-0.013\pm 0.005$ &SMARTS R \\ 
$3078.627$ &$  0.14$ &$\phantom{-} 0.082\pm 0.024$ &$\phantom{-} 1.508\pm 0.060$ &$\phantom{-} 0.027\pm 0.009$ &SMARTS J \\ 
$3078.631$ &$  3.41$ &$-0.063\pm 0.010$ &$\phantom{-} 1.393\pm 0.012$ &$\phantom{-} 0.049\pm 0.004$ &SMARTS R \\ 
$3090.616$ &$  3.62$ &$-0.069\pm 0.010$ &$\phantom{-} 1.384\pm 0.012$ &$\phantom{-} 0.048\pm 0.004$ &SMARTS R \\ 
$3105.529$ &$  0.13$ &$\phantom{-} 0.077\pm 0.041$ &$\phantom{-} 1.717\pm 0.138$ &$\phantom{-} 0.020\pm 0.009$ &SMARTS J \\ 
$3105.532$ &$  1.99$ &$-0.080\pm 0.010$ &$\phantom{-} 1.355\pm 0.013$ &$\phantom{-} 0.033\pm 0.005$ &SMARTS R \\ 
$3111.521$ &$  1.34$ &$-0.090\pm 0.011$ &$\phantom{-} 1.382\pm 0.014$ &$\phantom{-} 0.024\pm 0.005$ &SMARTS R \\ 
$3111.522$ &$  0.13$ &$\phantom{-} 0.070\pm 0.072$ &$\phantom{-} 1.653\pm 0.239$ &$\phantom{-} 0.012\pm 0.010$ &SMARTS J \\ 
$3121.535$ &$  1.00$ &$-0.061\pm 0.011$ &$\phantom{-} 1.392\pm 0.015$ &$\phantom{-} 0.017\pm 0.005$ &SMARTS R \\ 
$3122.520$ &$  0.74$ &$-0.054\pm 0.013$ &$\phantom{-} 1.314\pm 0.033$ &$\phantom{-} 0.001\pm 0.006$ &SMARTS R \\ 
$3123.493$ &$  0.19$ &$\phantom{-} 0.154\pm 0.040$ &$\phantom{-} 1.575\pm 0.115$ &$\phantom{-} 0.021\pm 0.009$ &SMARTS J \\ 
$3123.496$ &$  1.15$ &$-0.067\pm 0.010$ &$\phantom{-} 1.355\pm 0.013$ &$\phantom{-} 0.030\pm 0.005$ &SMARTS R \\ 
$3129.583$ &$  0.12$ &$\phantom{-} 0.130\pm 0.030$ &$\phantom{-} 1.569\pm 0.081$ &$\phantom{-} 0.026\pm 0.009$ &SMARTS J \\ 
$3129.584$ &$  1.74$ &$-0.014\pm 0.011$ &$\phantom{-} 1.393\pm 0.015$ &$\phantom{-} 0.013\pm 0.005$ &SMARTS R \\ 
$3137.523$ &$  0.12$ &$\phantom{-} 0.120\pm 0.053$ &$\phantom{-} 1.528\pm 0.154$ &$\phantom{-} 0.016\pm 0.010$ &SMARTS J \\ 
$3137.524$ &$  1.01$ &$-0.040\pm 0.011$ &$\phantom{-} 1.362\pm 0.014$ &$\phantom{-} 0.026\pm 0.005$ &SMARTS R \\ 
$3145.530$ &$  0.11$ &$\phantom{-} 0.084\pm 0.033$ &$\phantom{-} 1.483\pm 0.088$ &$\phantom{-} 0.022\pm 0.009$ &SMARTS J \\ 
$3145.533$ &$  1.07$ &$-0.056\pm 0.010$ &$\phantom{-} 1.370\pm 0.013$ &$\phantom{-} 0.037\pm 0.005$ &SMARTS R \\ 
$3152.459$ &$  1.94$ &$-0.022\pm 0.010$ &$\phantom{-} 1.361\pm 0.012$ &$\phantom{-} 0.041\pm 0.005$ &SMARTS R \\ 
$3159.494$ &$  0.15$ &$\phantom{-} 0.125\pm 0.028$ &$\phantom{-} 1.540\pm 0.071$ &$\phantom{-} 0.025\pm 0.009$ &SMARTS J \\ 
$3159.497$ &$  2.06$ &$\phantom{-} 0.021\pm 0.011$ &$\phantom{-} 1.372\pm 0.014$ &$\phantom{-} 0.019\pm 0.005$ &SMARTS R \\ 
$3165.454$ &$  0.24$ &$\phantom{-} 0.174\pm 0.040$ &$\phantom{-} 1.477\pm 0.103$ &$\phantom{-} 0.019\pm 0.009$ &SMARTS J \\ 
$3165.457$ &$  2.05$ &$\phantom{-} 0.045\pm 0.010$ &$\phantom{-} 1.371\pm 0.012$ &$\phantom{-} 0.039\pm 0.005$ &SMARTS R \\ 
$3174.443$ &$  0.22$ &$\phantom{-} 0.163\pm 0.050$ &$\phantom{-} 1.362\pm 0.121$ &$\phantom{-} 0.018\pm 0.010$ &SMARTS J \\ 
$3174.447$ &$  0.52$ &$\phantom{-} 0.032\pm 0.011$ &$\phantom{-} 1.353\pm 0.018$ &$-0.023\pm 0.005$ &SMARTS R \\ 
$3182.527$ &$  0.08$ &$\phantom{-} 0.200\pm 0.047$ &$\phantom{-} 1.528\pm 0.122$ &$-0.119\pm 0.011$ &SMARTS J \\ 
$3182.530$ &$  0.64$ &$\phantom{-} 0.033\pm 0.011$ &$\phantom{-} 1.348\pm 0.016$ &$-0.007\pm 0.005$ &SMARTS R \\ 
$3188.462$ &$  0.33$ &$\phantom{-} 0.204\pm 0.046$ &$\phantom{-} 1.674\pm 0.140$ &$\phantom{-} 0.018\pm 0.009$ &SMARTS J \\ 
$3188.465$ &$  0.67$ &$\phantom{-} 0.011\pm 0.011$ &$\phantom{-} 1.349\pm 0.017$ &$-0.013\pm 0.005$ &SMARTS R \\ 
$3196.492$ &$  0.09$ &$\phantom{-} 0.206\pm 0.061$ &$\phantom{-} 1.490\pm 0.155$ &$-0.119\pm 0.011$ &SMARTS J \\ 
$3196.493$ &$  0.69$ &$\phantom{-} 0.024\pm 0.011$ &$\phantom{-} 1.370\pm 0.014$ &$\phantom{-} 0.022\pm 0.005$ &SMARTS R \\ 
$3197.515$ &$  0.21$ &$\phantom{-} 0.118\pm 0.041$ &$\phantom{-} 1.404\pm 0.101$ &$-0.119\pm 0.011$ &SMARTS J \\ 
$3197.519$ &$  2.08$ &$\phantom{-} 0.034\pm 0.010$ &$\phantom{-} 1.375\pm 0.013$ &$\phantom{-} 0.033\pm 0.005$ &SMARTS R \\ 
$3206.470$ &$  1.86$ &$\phantom{-} 0.054\pm 0.011$ &$\phantom{-} 1.366\pm 0.013$ &$\phantom{-} 0.029\pm 0.005$ &SMARTS R \\ 
$3218.464$ &$  0.11$ &$\phantom{-} 0.155\pm 0.070$ &$\phantom{-} 1.338\pm 0.172$ &$\phantom{-} 0.014\pm 0.010$ &SMARTS J \\ 
$3218.468$ &$  0.41$ &$\phantom{-} 0.035\pm 0.014$ &$\phantom{-} 1.330\pm 0.029$ &$-0.070\pm 0.006$ &SMARTS R \\ 
$3315.853$ &$  0.21$ &$\phantom{-} 0.118\pm 0.071$ &$\phantom{-} 1.668\pm 0.160$ &$\phantom{-} 0.515\pm 0.021$ &SMARTS J \\ 
$3315.857$ &$  0.57$ &$\phantom{-} 0.024\pm 0.012$ &$\phantom{-} 1.317\pm 0.019$ &$-0.037\pm 0.005$ &SMARTS R \\ 
$3327.843$ &$  0.12$ &$\phantom{-} 0.124\pm 0.027$ &$\phantom{-} 1.488\pm 0.070$ &$\phantom{-} 0.030\pm 0.009$ &SMARTS J \\ 
$3327.845$ &$  1.28$ &$\phantom{-} 0.037\pm 0.011$ &$\phantom{-} 1.302\pm 0.013$ &$\phantom{-} 0.027\pm 0.005$ &SMARTS R \\ 
$3327.857$ &$  0.46$ &$-0.726\pm 0.015$ &$\phantom{-} 0.450\pm 0.027$ &$-0.062\pm 0.007$ &SMARTS B \\ 
$3334.816$ &$  0.66$ &$\phantom{-} 0.043\pm 0.013$ &$\phantom{-} 1.361\pm 0.025$ &$-0.057\pm 0.005$ &SMARTS R \\ 
$3335.849$ &$  0.18$ &$\phantom{-} 0.120\pm 0.031$ &$\phantom{-} 1.541\pm 0.083$ &$\phantom{-} 0.022\pm 0.009$ &SMARTS J \\ 
$3335.852$ &$  0.81$ &$\phantom{-} 0.035\pm 0.011$ &$\phantom{-} 1.282\pm 0.016$ &$-0.011\pm 0.005$ &SMARTS R \\ 
$3346.771$ &$  0.18$ &$\phantom{-} 0.225\pm 0.064$ &$\phantom{-} 1.812\pm 0.223$ &$\phantom{-} 0.013\pm 0.010$ &SMARTS J \\ 
$3346.774$ &$  0.67$ &$\phantom{-} 0.051\pm 0.014$ &$\phantom{-} 1.249\pm 0.026$ &$-0.095\pm 0.006$ &SMARTS R \\ 
$3346.785$ &$  0.55$ &$-0.755\pm 0.011$ &$\phantom{-} 0.402\pm 0.015$ &$-0.019\pm 0.005$ &SMARTS B \\ 
$3354.771$ &$  0.09$ &$\phantom{-} 0.175\pm 0.076$ &$\phantom{-} 1.412\pm 0.195$ &$\phantom{-} 0.014\pm 0.010$ &SMARTS J \\ 
$3354.773$ &$  0.69$ &$\phantom{-} 0.024\pm 0.012$ &$\phantom{-} 1.280\pm 0.017$ &$-0.018\pm 0.005$ &SMARTS R \\ 
$3354.785$ &$  0.55$ &$-0.770\pm 0.014$ &$\phantom{-} 0.425\pm 0.021$ &$-0.057\pm 0.006$ &SMARTS B \\ 
$3358.814$ &$  1.25$ &$-0.766\pm 0.011$ &$\phantom{-} 0.389\pm 0.012$ &$\phantom{-} 0.045\pm 0.005$ &SMARTS B \\ 
$3361.738$ &$  0.76$ &$\phantom{-} 0.026\pm 0.011$ &$\phantom{-} 1.257\pm 0.015$ &$\phantom{-} 0.003\pm 0.005$ &SMARTS R \\ 
$3370.777$ &$  0.13$ &$\phantom{-} 0.133\pm 0.032$ &$\phantom{-} 1.542\pm 0.088$ &$\phantom{-} 0.024\pm 0.009$ &SMARTS J \\ 
$3370.780$ &$  1.90$ &$\phantom{-} 0.022\pm 0.011$ &$\phantom{-} 1.264\pm 0.013$ &$\phantom{-} 0.021\pm 0.005$ &SMARTS R \\ 
$3377.744$ &$  1.93$ &$\phantom{-} 0.038\pm 0.010$ &$\phantom{-} 1.266\pm 0.012$ &$\phantom{-} 0.038\pm 0.005$ &SMARTS R \\ 
$3384.692$ &$  1.08$ &$\phantom{-} 0.048\pm 0.012$ &$\phantom{-} 1.270\pm 0.018$ &$-0.025\pm 0.005$ &SMARTS R \\ 
$3384.704$ &$  1.21$ &$-0.736\pm 0.012$ &$\phantom{-} 0.450\pm 0.015$ &$-0.017\pm 0.005$ &SMARTS B \\ 
$3397.666$ &$  0.22$ &$\phantom{-} 0.173\pm 0.041$ &$\phantom{-} 1.682\pm 0.121$ &$-0.119\pm 0.011$ &SMARTS J \\ 
$3397.668$ &$  0.52$ &$\phantom{-} 0.040\pm 0.012$ &$\phantom{-} 1.249\pm 0.018$ &$-0.031\pm 0.005$ &SMARTS R \\ 
$3406.804$ &$  0.17$ &$\phantom{-} 0.110\pm 0.027$ &$\phantom{-} 1.337\pm 0.056$ &$-0.119\pm 0.011$ &SMARTS J \\ 
$3406.808$ &$  2.30$ &$\phantom{-} 0.025\pm 0.010$ &$\phantom{-} 1.231\pm 0.012$ &$\phantom{-} 0.039\pm 0.005$ &SMARTS R \\ 
$3413.764$ &$  0.14$ &$\phantom{-} 0.130\pm 0.025$ &$\phantom{-} 1.459\pm 0.062$ &$\phantom{-} 0.033\pm 0.009$ &SMARTS J \\ 
$3413.767$ &$  3.09$ &$\phantom{-} 0.012\pm 0.010$ &$\phantom{-} 1.191\pm 0.012$ &$\phantom{-} 0.039\pm 0.005$ &SMARTS R \\ 
$3413.779$ &$  2.14$ &$-0.773\pm 0.011$ &$\phantom{-} 0.292\pm 0.012$ &$\phantom{-} 0.043\pm 0.005$ &SMARTS B \\ 
$3424.707$ &$  0.16$ &$\phantom{-} 0.107\pm 0.028$ &$\phantom{-} 1.438\pm 0.065$ &$-0.119\pm 0.011$ &SMARTS J \\ 
$3424.711$ &$  0.66$ &$\phantom{-} 0.004\pm 0.011$ &$\phantom{-} 1.214\pm 0.016$ &$-0.014\pm 0.005$ &SMARTS R \\ 
$3431.770$ &$  0.25$ &$\phantom{-} 0.171\pm 0.027$ &$\phantom{-} 1.290\pm 0.055$ &$\phantom{-} 0.026\pm 0.009$ &SMARTS J \\ 
$3431.773$ &$  1.75$ &$\phantom{-} 0.010\pm 0.011$ &$\phantom{-} 1.172\pm 0.013$ &$\phantom{-} 0.027\pm 0.005$ &SMARTS R \\ 
$3439.735$ &$  0.14$ &$\phantom{-} 0.147\pm 0.029$ &$\phantom{-} 1.336\pm 0.065$ &$\phantom{-} 0.027\pm 0.009$ &SMARTS J \\ 
$3439.739$ &$  1.67$ &$\phantom{-} 0.009\pm 0.011$ &$\phantom{-} 1.146\pm 0.012$ &$\phantom{-} 0.028\pm 0.005$ &SMARTS R \\ 
$3439.751$ &$  1.03$ &$-0.767\pm 0.011$ &$\phantom{-} 0.242\pm 0.012$ &$\phantom{-} 0.025\pm 0.005$ &SMARTS B \\ 
$3447.616$ &$  0.14$ &$\phantom{-} 0.124\pm 0.030$ &$\phantom{-} 1.438\pm 0.069$ &$-0.119\pm 0.011$ &SMARTS J \\ 
$3447.619$ &$  1.57$ &$-0.014\pm 0.010$ &$\phantom{-} 1.162\pm 0.012$ &$\phantom{-} 0.033\pm 0.005$ &SMARTS R \\ 
$3458.637$ &$  0.15$ &$\phantom{-} 0.088\pm 0.027$ &$\phantom{-} 1.424\pm 0.066$ &$\phantom{-} 0.029\pm 0.009$ &SMARTS J \\ 
$3458.640$ &$  1.73$ &$-0.028\pm 0.011$ &$\phantom{-} 1.146\pm 0.013$ &$\phantom{-} 0.017\pm 0.005$ &SMARTS R \\ 
$3465.602$ &$  0.10$ &$\phantom{-} 0.088\pm 0.037$ &$\phantom{-} 1.436\pm 0.090$ &$-0.119\pm 0.011$ &SMARTS J \\ 
$3465.604$ &$  1.33$ &$-0.038\pm 0.011$ &$\phantom{-} 1.142\pm 0.012$ &$\phantom{-} 0.031\pm 0.005$ &SMARTS R \\ 
$3471.548$ &$  0.50$ &$\phantom{-} 0.088\pm 0.018$ &$\phantom{-} 1.437\pm 0.043$ &$\phantom{-} 0.045\pm 0.009$ &SMARTS J \\ 
$3471.552$ &$  0.92$ &$-0.037\pm 0.011$ &$\phantom{-} 1.169\pm 0.014$ &$\phantom{-} 0.012\pm 0.005$ &SMARTS R \\ 
$3471.564$ &$  0.61$ &$-0.820\pm 0.011$ &$\phantom{-} 0.284\pm 0.014$ &$\phantom{-} 0.014\pm 0.005$ &SMARTS B \\ 
$3472.557$ &$  0.91$ &$-0.037\pm 0.011$ &$\phantom{-} 1.160\pm 0.014$ &$\phantom{-} 0.011\pm 0.005$ &SMARTS R \\ 
$3472.570$ &$  0.49$ &$-0.826\pm 0.011$ &$\phantom{-} 0.272\pm 0.013$ &$\phantom{-} 0.022\pm 0.005$ &SMARTS B \\ 
$3475.599$ &$  0.14$ &$\phantom{-} 0.048\pm 0.020$ &$\phantom{-} 1.385\pm 0.043$ &$\phantom{-} 0.039\pm 0.009$ &SMARTS J \\ 
$3475.602$ &$  2.42$ &$-0.034\pm 0.010$ &$\phantom{-} 1.165\pm 0.012$ &$\phantom{-} 0.038\pm 0.005$ &SMARTS R \\ 
$3475.613$ &$  0.90$ &$-0.837\pm 0.010$ &$\phantom{-} 0.268\pm 0.012$ &$\phantom{-} 0.050\pm 0.005$ &SMARTS B \\ 
$3482.537$ &$  0.14$ &$\phantom{-} 0.067\pm 0.040$ &$\phantom{-} 1.486\pm 0.115$ &$\phantom{-} 0.020\pm 0.009$ &SMARTS J \\ 
$3482.543$ &$  0.71$ &$-0.005\pm 0.015$ &$\phantom{-} 1.172\pm 0.029$ &$-0.103\pm 0.006$ &SMARTS R \\ 
$3492.565$ &$  0.22$ &$\phantom{-} 0.071\pm 0.020$ &$\phantom{-} 1.386\pm 0.043$ &$\phantom{-} 0.039\pm 0.009$ &SMARTS J \\ 
$3492.567$ &$  2.82$ &$-0.023\pm 0.010$ &$\phantom{-} 1.180\pm 0.012$ &$\phantom{-} 0.040\pm 0.005$ &SMARTS R \\ 
$3492.581$ &$  2.10$ &$-0.815\pm 0.011$ &$\phantom{-} 0.291\pm 0.012$ &$\phantom{-} 0.047\pm 0.005$ &SMARTS B \\ 
$3504.532$ &$  0.66$ &$-0.024\pm 0.011$ &$\phantom{-} 1.160\pm 0.014$ &$\phantom{-} 0.007\pm 0.005$ &SMARTS R \\ 
$3515.620$ &$  0.47$ &$-0.037\pm 0.013$ &$\phantom{-} 1.117\pm 0.024$ &$-0.068\pm 0.006$ &SMARTS R \\ 
$3517.567$ &$  0.10$ &$\phantom{-} 0.037\pm 0.044$ &$\phantom{-} 1.317\pm 0.109$ &$\phantom{-} 0.017\pm 0.009$ &SMARTS J \\ 
$3517.570$ &$  0.68$ &$-0.046\pm 0.011$ &$\phantom{-} 1.172\pm 0.015$ &$-0.006\pm 0.005$ &SMARTS R \\ 
$3525.508$ &$  0.19$ &$\phantom{-} 0.048\pm 0.024$ &$\phantom{-} 1.306\pm 0.052$ &$\phantom{-} 0.032\pm 0.009$ &SMARTS J \\ 
$3525.511$ &$  1.84$ &$-0.065\pm 0.010$ &$\phantom{-} 1.164\pm 0.012$ &$\phantom{-} 0.033\pm 0.005$ &SMARTS R \\ 
$3525.524$ &$  1.24$ &$-0.857\pm 0.011$ &$\phantom{-} 0.282\pm 0.012$ &$\phantom{-} 0.037\pm 0.005$ &SMARTS B \\ 
$3538.557$ &$  0.60$ &$-0.084\pm 0.012$ &$\phantom{-} 1.108\pm 0.019$ &$-0.051\pm 0.005$ &SMARTS R \\ 
$3551.520$ &$  0.68$ &$-0.891\pm 0.014$ &$\phantom{-} 0.335\pm 0.023$ &$-0.076\pm 0.007$ &SMARTS B \\ 
$3551.538$ &$  0.43$ &$-0.066\pm 0.024$ &$\phantom{-} 1.150\pm 0.057$ &$-0.097\pm 0.008$ &SMARTS R \\ 
$3551.551$ &$  0.66$ &$\phantom{-} 0.023\pm 0.027$ &$\phantom{-} 1.322\pm 0.069$ &$-0.072\pm 0.008$ &SMARTS I \\ 
$3563.475$ &$  2.56$ &$\phantom{-} 0.030\pm 0.010$ &$\phantom{-} 1.291\pm 0.013$ &$\phantom{-} 0.040\pm 0.005$ &SMARTS I \\ 
$3571.492$ &$  0.46$ &$-0.191\pm 0.031$ &$\phantom{-} 0.973\pm 0.078$ &$-0.060\pm 0.009$ &SMARTS R \\ 
$3576.461$ &$  0.21$ &$-0.001\pm 0.040$ &$\phantom{-} 1.483\pm 0.113$ &$\phantom{-} 0.015\pm 0.009$ &SMARTS J \\ 
$3576.463$ &$  1.04$ &$-0.155\pm 0.011$ &$\phantom{-} 1.161\pm 0.014$ &$\phantom{-} 0.006\pm 0.005$ &SMARTS R \\ 
$3584.453$ &$  0.49$ &$-0.158\pm 0.024$ &$\phantom{-} 1.108\pm 0.065$ &$-0.047\pm 0.008$ &SMARTS R \\ 
$3584.464$ &$  0.60$ &$-0.971\pm 0.014$ &$\phantom{-} 0.221\pm 0.027$ &$-0.050\pm 0.007$ &SMARTS B \\ 
$3690.852$ &$  0.66$ &$-0.189\pm 0.011$ &$\phantom{-} 1.100\pm 0.016$ &$-0.030\pm 0.005$ &SMARTS R \\ 
$3699.820$ &$  0.77$ &$-0.184\pm 0.011$ &$\phantom{-} 1.095\pm 0.014$ &$\phantom{-} 0.005\pm 0.005$ &SMARTS R \\ 
$3699.832$ &$  0.63$ &$-0.958\pm 0.011$ &$\phantom{-} 0.216\pm 0.014$ &$-0.017\pm 0.005$ &SMARTS B \\ 
$3708.804$ &$  1.44$ &$-0.178\pm 0.010$ &$\phantom{-} 1.059\pm 0.012$ &$\phantom{-} 0.028\pm 0.005$ &SMARTS R \\ 
$3708.816$ &$  0.92$ &$-0.939\pm 0.011$ &$\phantom{-} 0.196\pm 0.012$ &$\phantom{-} 0.021\pm 0.005$ &SMARTS B \\ 
$3715.827$ &$  0.63$ &$-0.028\pm 0.014$ &$\phantom{-} 1.283\pm 0.025$ &$\phantom{-} 0.073\pm 0.008$ &SMARTS J \\ 
$3715.828$ &$  1.48$ &$-0.165\pm 0.010$ &$\phantom{-} 1.046\pm 0.012$ &$\phantom{-} 0.036\pm 0.005$ &SMARTS R \\ 
$3722.766$ &$  0.64$ &$\phantom{-} 0.032\pm 0.021$ &$\phantom{-} 1.471\pm 0.051$ &$\phantom{-} 0.032\pm 0.009$ &SMARTS J \\ 
$3722.769$ &$  0.57$ &$-0.166\pm 0.012$ &$\phantom{-} 1.044\pm 0.017$ &$-0.040\pm 0.005$ &SMARTS R \\ 
$3724.763$ &$  0.81$ &$-0.912\pm 0.013$ &$\phantom{-} 0.226\pm 0.018$ &$-0.053\pm 0.006$ &SMARTS B \\ 
$3724.775$ &$  1.51$ &$-0.158\pm 0.011$ &$\phantom{-} 1.053\pm 0.014$ &$\phantom{-} 0.000\pm 0.005$ &SMARTS R \\ 
$3724.787$ &$  1.64$ &$-0.025\pm 0.011$ &$\phantom{-} 1.196\pm 0.016$ &$\phantom{-} 0.004\pm 0.005$ &SMARTS I \\ 
$3725.813$ &$  0.97$ &$-0.033\pm 0.015$ &$\phantom{-} 1.341\pm 0.027$ &$\phantom{-} 0.070\pm 0.009$ &SMARTS J \\ 
$3747.735$ &$  1.37$ &$-0.058\pm 0.011$ &$\phantom{-} 1.185\pm 0.015$ &$\phantom{-} 0.010\pm 0.005$ &SMARTS I \\ 
$3747.749$ &$  0.44$ &$-0.917\pm 0.012$ &$\phantom{-} 0.202\pm 0.015$ &$-0.038\pm 0.005$ &SMARTS B \\ 
$3747.761$ &$  1.06$ &$-0.172\pm 0.011$ &$\phantom{-} 1.051\pm 0.013$ &$\phantom{-} 0.016\pm 0.005$ &SMARTS R \\ 
$3748.742$ &$  0.74$ &$-0.006\pm 0.016$ &$\phantom{-} 1.385\pm 0.033$ &$\phantom{-} 0.043\pm 0.009$ &SMARTS J \\ 
$3749.760$ &$  0.04$ &$\phantom{-}0.048\pm 0.043$ &$\phantom{-} 1.374\pm 0.089$ & $\equiv 0$ &SOAR J \\ 
$3749.769$ &$  0.14$ &$\phantom{-}0.008\pm 0.031$ &$\phantom{-} 1.307\pm 0.047$ & $\equiv 0$ &SOAR H \\ 
$3749.778$ &$  0.40$ &$-0.675\pm 0.030$ &$\phantom{-} 0.546\pm 0.038$ & $\equiv 0$ &SOAR $\rm K_s$ \\ 
$3754.774$ &$  0.77$ &$-0.176\pm 0.012$ &$\phantom{-} 1.030\pm 0.019$ &$-0.062\pm 0.005$ &SMARTS R \\ 
$3760.737$ &$  0.21$ &$\phantom{-} 0.001\pm 0.022$ &$\phantom{-} 1.415\pm 0.047$ &$-0.119\pm 0.011$ &SMARTS J \\ 
$3760.740$ &$  1.18$ &$-0.891\pm 0.011$ &$\phantom{-} 0.173\pm 0.012$ &$\phantom{-} 0.037\pm 0.005$ &SMARTS B \\ 
$3760.753$ &$  1.34$ &$-0.163\pm 0.010$ &$\phantom{-} 1.045\pm 0.012$ &$\phantom{-} 0.030\pm 0.005$ &SMARTS R \\ 
$3760.765$ &$  2.11$ &$-0.059\pm 0.011$ &$\phantom{-} 1.168\pm 0.014$ &$\phantom{-} 0.017\pm 0.005$ &SMARTS I \\ 
$3770.750$ &$  0.27$ &$-0.009\pm 0.019$ &$\phantom{-} 1.326\pm 0.036$ &$-0.118\pm 0.011$ &SMARTS J \\ 
$3770.751$ &$  1.04$ &$-0.867\pm 0.011$ &$\phantom{-} 0.167\pm 0.012$ &$\phantom{-} 0.032\pm 0.005$ &SMARTS B \\ 
$3770.764$ &$  1.67$ &$-0.155\pm 0.010$ &$\phantom{-} 1.018\pm 0.012$ &$\phantom{-} 0.029\pm 0.005$ &SMARTS R \\ 
$3776.634$ &$  0.24$ &$-0.036\pm 0.020$ &$\phantom{-} 1.245\pm 0.038$ &$-0.119\pm 0.011$ &SMARTS J \\ 
$3776.637$ &$  0.83$ &$-0.162\pm 0.011$ &$\phantom{-} 1.012\pm 0.014$ &$-0.004\pm 0.005$ &SMARTS R \\ 
$3784.704$ &$  0.18$ &$-0.005\pm 0.022$ &$\phantom{-} 1.378\pm 0.045$ &$-0.119\pm 0.011$ &SMARTS J \\ 
$3784.707$ &$  0.73$ &$-0.141\pm 0.011$ &$\phantom{-} 1.019\pm 0.014$ &$-0.010\pm 0.005$ &SMARTS R \\ 
$3790.694$ &$  0.15$ &$-0.013\pm 0.020$ &$\phantom{-} 1.294\pm 0.041$ &$\phantom{-} 0.039\pm 0.009$ &SMARTS J \\ 
$3790.696$ &$  1.17$ &$-0.886\pm 0.011$ &$\phantom{-} 0.118\pm 0.012$ &$\phantom{-} 0.034\pm 0.005$ &SMARTS B \\ 
$3790.708$ &$  1.75$ &$-0.154\pm 0.010$ &$\phantom{-} 1.009\pm 0.012$ &$\phantom{-} 0.036\pm 0.005$ &SMARTS R \\ 
$3797.651$ &$  0.17$ &$-0.014\pm 0.021$ &$\phantom{-} 1.299\pm 0.043$ &$\phantom{-} 0.033\pm 0.009$ &SMARTS J \\ 
$3797.653$ &$  0.91$ &$-0.916\pm 0.011$ &$\phantom{-} 0.092\pm 0.012$ &$\phantom{-} 0.041\pm 0.005$ &SMARTS B \\ 
$3797.666$ &$  1.85$ &$-0.173\pm 0.010$ &$\phantom{-} 0.987\pm 0.012$ &$\phantom{-} 0.038\pm 0.005$ &SMARTS R \\ 
$3804.663$ &$  0.17$ &$-0.011\pm 0.019$ &$\phantom{-} 1.223\pm 0.038$ &$\phantom{-} 0.039\pm 0.009$ &SMARTS J \\ 
$3804.665$ &$  0.65$ &$-0.941\pm 0.011$ &$\phantom{-} 0.057\pm 0.013$ &$-0.011\pm 0.005$ &SMARTS B \\ 
$3806.677$ &$  0.26$ &$\phantom{-} 0.031\pm 0.020$ &$\phantom{-} 1.361\pm 0.045$ &$\phantom{-} 0.036\pm 0.009$ &SMARTS J \\ 
$3806.680$ &$  1.31$ &$-0.174\pm 0.011$ &$\phantom{-} 0.956\pm 0.013$ &$\phantom{-} 0.008\pm 0.005$ &SMARTS R \\ 
$3808.663$ &$  0.60$ &$-0.923\pm 0.016$ &$\phantom{-} 0.051\pm 0.024$ &$-0.069\pm 0.007$ &SMARTS B \\ 
$3809.656$ &$  0.22$ &$-0.067\pm 0.025$ &$\phantom{-} 1.198\pm 0.054$ &$\phantom{-} 0.022\pm 0.009$ &SMARTS J \\ 
$3809.660$ &$  0.98$ &$-0.185\pm 0.015$ &$\phantom{-} 0.917\pm 0.027$ &$-0.124\pm 0.006$ &SMARTS R \\ 
$3819.586$ &$  0.16$ &$-0.058\pm 0.021$ &$\phantom{-} 1.288\pm 0.045$ &$\phantom{-} 0.031\pm 0.009$ &SMARTS J \\ 
$3821.605$ &$  0.19$ &$-0.067\pm 0.022$ &$\phantom{-} 1.326\pm 0.048$ &$\phantom{-} 0.029\pm 0.009$ &SMARTS J \\ 
$3821.608$ &$  1.76$ &$-0.988\pm 0.011$ &$\phantom{-} 0.054\pm 0.012$ &$\phantom{-} 0.037\pm 0.005$ &SMARTS B \\ 
$3826.569$ &$  0.18$ &$-0.060\pm 0.041$ &$\phantom{-} 1.361\pm 0.100$ &$-0.119\pm 0.011$ &SMARTS J \\ 
$3832.622$ &$  0.13$ &$-0.017\pm 0.025$ &$\phantom{-} 1.273\pm 0.057$ &$\phantom{-} 0.027\pm 0.009$ &SMARTS J \\ 
$3832.624$ &$  0.65$ &$-0.979\pm 0.011$ &$-0.005\pm 0.013$ &$-0.031\pm 0.005$ &SMARTS B \\ 
$3833.598$ &$  0.13$ &$-0.041\pm 0.028$ &$\phantom{-} 1.281\pm 0.067$ &$\phantom{-} 0.021\pm 0.009$ &SMARTS J \\ 
$3833.602$ &$  0.83$ &$-0.232\pm 0.011$ &$\phantom{-} 0.913\pm 0.013$ &$\phantom{-} 0.007\pm 0.005$ &SMARTS R \\ 
$3839.588$ &$  0.43$ &$-0.242\pm 0.014$ &$\phantom{-} 0.874\pm 0.023$ &$-0.083\pm 0.006$ &SMARTS R \\ 
$3846.572$ &$  0.17$ &$-0.074\pm 0.028$ &$\phantom{-} 1.360\pm 0.070$ &$\phantom{-} 0.019\pm 0.009$ &SMARTS J \\ 
$3846.575$ &$  0.71$ &$-0.989\pm 0.011$ &$-0.027\pm 0.012$ &$\phantom{-} 0.017\pm 0.005$ &SMARTS B \\ 
$3847.539$ &$  0.20$ &$-0.036\pm 0.030$ &$\phantom{-} 0.929\pm 0.053$ &$\phantom{-} 0.021\pm 0.009$ &SMARTS J \\ 
$3847.542$ &$  0.93$ &$-0.247\pm 0.011$ &$\phantom{-} 0.894\pm 0.013$ &$\phantom{-} 0.010\pm 0.005$ &SMARTS R \\ 
$3854.638$ &$  0.86$ &$-0.250\pm 0.011$ &$\phantom{-} 0.899\pm 0.013$ &$-0.006\pm 0.005$ &SMARTS R \\ 
$3863.464$ &$  0.20$ &$-0.057\pm 0.027$ &$\phantom{-} 1.228\pm 0.060$ &$\phantom{-} 0.022\pm 0.009$ &SMARTS J \\ 
$3863.467$ &$  0.62$ &$-0.254\pm 0.012$ &$\phantom{-} 0.844\pm 0.017$ &$-0.072\pm 0.005$ &SMARTS R \\ 
$3870.483$ &$  0.18$ &$-0.079\pm 0.028$ &$\phantom{-} 1.251\pm 0.062$ &$-0.119\pm 0.011$ &SMARTS J \\ 
$3870.486$ &$  0.54$ &$-0.265\pm 0.011$ &$\phantom{-} 0.873\pm 0.014$ &$-0.015\pm 0.005$ &SMARTS R \\ 
$3877.546$ &$  1.64$ &$-1.031\pm 0.011$ &$-0.088\pm 0.012$ &$\phantom{-} 0.028\pm 0.005$ &SMARTS B \\ 
$3877.557$ &$  2.08$ &$-0.278\pm 0.010$ &$\phantom{-} 0.876\pm 0.012$ &$\phantom{-} 0.028\pm 0.005$ &SMARTS R \\ 
$3883.447$ &$  0.22$ &$-0.102\pm 0.028$ &$\phantom{-} 1.271\pm 0.061$ &$-0.119\pm 0.011$ &SMARTS J \\ 
$3883.451$ &$  0.44$ &$-1.028\pm 0.012$ &$-0.043\pm 0.015$ &$-0.054\pm 0.006$ &SMARTS B \\ 
$3883.463$ &$  1.46$ &$-0.279\pm 0.010$ &$\phantom{-} 0.874\pm 0.012$ &$\phantom{-} 0.028\pm 0.005$ &SMARTS R \\ 
$3891.501$ &$  0.53$ &$-0.281\pm 0.011$ &$\phantom{-} 0.894\pm 0.015$ &$-0.034\pm 0.005$ &SMARTS R \\ 
$3898.500$ &$  3.03$ &$-0.097\pm 0.024$ &$\phantom{-} 0.910\pm 0.038$ &$-0.119\pm 0.011$ &SMARTS J \\ 
$3898.503$ &$  1.08$ &$-0.291\pm 0.011$ &$\phantom{-} 0.880\pm 0.013$ &$-0.004\pm 0.005$ &SMARTS R \\ 
$3904.444$ &$  0.28$ &$-0.093\pm 0.026$ &$\phantom{-} 1.136\pm 0.054$ &$\phantom{-} 0.025\pm 0.009$ &SMARTS J \\ 
$3904.448$ &$  0.87$ &$-0.304\pm 0.011$ &$\phantom{-} 0.865\pm 0.015$ &$-0.027\pm 0.005$ &SMARTS R \\ 
$3915.482$ &$  0.19$ &$-0.085\pm 0.027$ &$\phantom{-} 1.197\pm 0.056$ &$-0.119\pm 0.011$ &SMARTS J \\ 
$3915.485$ &$  1.16$ &$-1.023\pm 0.011$ &$-0.032\pm 0.012$ &$\phantom{-} 0.024\pm 0.005$ &SMARTS B \\ 
$3915.496$ &$  1.37$ &$-0.288\pm 0.010$ &$\phantom{-} 0.888\pm 0.012$ &$\phantom{-} 0.025\pm 0.005$ &SMARTS R \\ 
$3925.461$ &$  0.87$ &$-0.302\pm 0.011$ &$\phantom{-} 0.873\pm 0.016$ &$-0.041\pm 0.005$ &SMARTS R \\ 
$3932.475$ &$  0.19$ &$-0.066\pm 0.113$ &$\phantom{-} 0.808\pm 0.200$ &$\phantom{-} 0.012\pm 0.010$ &SMARTS J \\ 
$3932.478$ &$  0.47$ &$-0.287\pm 0.013$ &$\phantom{-} 0.842\pm 0.020$ &$-0.079\pm 0.006$ &SMARTS R \\ 
$3944.470$ &$  0.96$ &$-0.324\pm 0.011$ &$\phantom{-} 0.840\pm 0.015$ &$-0.034\pm 0.005$ &SMARTS R \\ 
$3946.460$ &$  0.54$ &$-0.354\pm 0.018$ &$\phantom{-} 0.832\pm 0.038$ &$-0.097\pm 0.007$ &SMARTS R \\ 

%% file: tab4.tex
AIC & $N_s=40$ & $153.0^{+2.4}_{-2.2}$ \\
BIC & $N_s=40$ & $152.2^{+2.8}_{-3.0}$ \\
AIC & $N_s=70$ & $152.6^{+1.8}_{-1.8}$ \\
BIC & $N_s=70$ & $150.6^{+1.6}_{-1.8}$ \\
F-Test & $N_s=40$ & $157.2^{+2.6}_{-2.6}$ \\

%% file: ms.bbl
\begin{thebibliography}{}

\bibitem[Agol, Jones, \& Blaes(2000)]{Agol00}
Agol, E., Jones, B., Blaes, O., 2000, \apj, 545, 657

\bibitem[Bar-Kana(1996)]{Barkana96}
Bar-Kana, R., 1996, \apj, 468, 17B

\bibitem[Barvainis(1987)]{Barvainis87}
Barvainis, R., 1987, \apj, 320, 537

\bibitem[Chiba et al.(2005)]{Chiba05}
Chiba, M., Minezaki, T., Kashikawa, N., Kataza, H., Inoue, K.T., 2005, \apj, 627, 53C

\bibitem[Courbin, Lidman \& Magain(1998)]{Courbin98}
Courbin, F., Lidman, C., \& Magain, P., 1998, \aap, 330, 57

\bibitem[Dalal \& Kochanek(2002)]{Dalal02}
Dalal, N., Kochanek, C.S., 2002, \apj, 572, 25

\bibitem[DePoy et al.(1993)]{DePoy93}
DePoy, D. L., Atwood, B., Byard, P. L., Frogel, J., \& O'Brien, T. P.,
1993, SPIE, 1946, 667

\bibitem[DePoy et al.(2003)]{DePoy03}
DePoy, D.L., Atwood, B., Belville, S.R., Brewer, D.F., Byard, P.L.,
Gould, A., Mason, J.A., O'Brien, T.P., Pappalardo, D.P., Pogge, R.W.,
Steinbrecher, D.P., \& Tiega, E.J., 2003, SPIE, 4841, 827

\bibitem[Falco et al.(1999)]{Falco99}
Falco, E. E., 1999, \apj, 523, 617F

\bibitem[Fazio et al.(2004)]{Fazio04}
Fazio, G. G., et al. 2004, \apjs, 154, 10

\bibitem[Freedman et al.(2001)]{Freedman01}
Freedman, W.L., et al., 2001, \apj, 553, 47

\bibitem[Gil-Merino, Wisotzki, \& Wambsganss(2002)]{Gil-Merino02}
Gil-Merino, R., Wisotzki, L., \& Wambsganss, J., 2002, \aap, 381, 428

\bibitem[Keeton(2001)]{Keeton01}
Keeton, C.R., 2001, astro-ph/0102340

\bibitem[Kochanek(2002)]{Kochanek02}
Kochanek, C.S., 2002, \apj, 587, 25


\bibitem[Kochanek et al.(2006)]{Kochanek06a}
Kochanek, C.S., Morgan, N.D., Falco, E.E., McLeod, B.A., Winn J.N.,
Dembicky J., \& Ketzeback, B., 2006, \apj, 640, 47


\bibitem[Leh\'ar et al.(2000)]{Lehar00}
Leh\'ar, J., et al., 2000, \apj, 536, 584

\bibitem[Mao \& Schneider(1998)]{Mao98}
Mao, S., Schneider, P., 1998, \mnras, 295, 587

\bibitem[Mu\~noz et al.(2004)]{Munoz04}
Mu\~noz, J. A., Falco, E. E., Kochanek, C. S., McLeod, B. A., \& Mediavilla, E.,
2004, \apj, 605, 614M

\bibitem[Ofek \& Maoz(2003)]{Ofek03}
Ofek, E.O. \& Maoz, D., 2003, \apj, 594, 101

\bibitem[Poindexter, Morgan \& Kochanek(2007)]{Poindexter07}
Poindexter, S., Morgan, N., \& Kochanek, C. S., 2007, work in progress

\bibitem[Refsdal(1964)]{Refsdal64}
Refsdal, S. 1964, \mnras, 182, 307

\bibitem[Remy et al.(1998)]{Remy98}
Remy, M., Claeskens, J.-F., Surdej, J., Hjorth, J., Refsdal, S., Wucknitz, O.,
S{\o}rensen, A.N., \& Grundahl, F., 1998, \na, 3, 379

\bibitem[Richards et al.(2004)]{Richards04}
Richards, G. T., Keeton, C.R., Pindor, B., et al., 2004, \apj, 610, 679R

\bibitem[Rusin \& Kochanek(2005)]{Rusin05}
Rusin, D., Kochanek, C.S., 2005, \apj, 623, 666

\bibitem[Schechter et al.(2003)]{Schechter03}
Schechter, P.L., et al. 2003, \apj, 584, 657

\bibitem[Treu et al.(2006)]{Treu06}
Treu, T., Koopmans, L.V., Bolton, A.S., Burles, S., \& Moustakas, L.A., 2006,
\apj, 640, 662

\bibitem[Wambsganss(2006)]{Wambsganss06}
Wambsganss, J, 2006, in Gravitational Lensing: Strong Weak and Micro, Saas-Fee
Advanced Course 33, G. Meylan, P. North, P. Jetzer, eds., (Springer: Berlin)
453, [astro-ph/0604278]

\bibitem[Wisotzki et al.(1993)]{Wisotzki93}
Wisotzki, L., K\"ohler, R., Kayser, R., \& Reimers, D., 1993, \aap, 278, L15

\bibitem[Wisotzki et al.(1998)]{Wisotzki98}
Wisotzki, L., Wucknitz, O., Lopez, S., \& S{\o}rensen, A.N., 1998, \aap, 339, L73

\bibitem[Wyrzykowski et al.(2003)]{Wyrzykowski03} 
Wyrzykowski, \L. et al., 2003 \actaa, 53, 229

\end{thebibliography}
